\documentclass[conference]{IEEEtran}

\usepackage[caption=false]{subfig}

\usepackage{hhline}

\usepackage{array}
\newcolumntype{P}[1]{>{\centering\arraybackslash}p{#1}}
\newcolumntype{M}[1]{>{\centering\arraybackslash}m{#1}}

\usepackage[pdftex]{graphicx}

\usepackage[linesnumbered,ruled,vlined]{algorithm2e}
\usepackage[dvipsnames,svgnames]{xcolor}
\usepackage{xspace}

\SetCommentSty{commentsty}
\SetKwComment{Comment}{$\triangleright$\ }{}
\makeatletter 
\g@addto@macro{\@algocf@init}{\SetKwInOut{Parameter}{Parameters}} 
\makeatother

\graphicspath{{fig/}}
\usepackage{amsmath}
\usepackage{amssymb} 
\DeclareGraphicsExtensions{.pdf,.jpeg,.png}
\usepackage{url}
\usepackage{import}
\usepackage{cleveref}

\definecolor{red2}{rgb}{0.84,0.21,0.14}
\definecolor{blue2}{rgb}{0.24,0.41,0.64}
\definecolor{green2}{rgb}{0.25,0.6,0.43}

\DeclareMathOperator*{\argmin}{argmin}
\DeclareMathOperator*{\argmax}{argmax}

\newcommand{\MP}[0]{\mathcal{P}}
\newcommand{\MV}[0]{\mathcal{V}}
\newcommand{\MA}[0]{\mathcal{A}}
\newcommand{\MX}[0]{\mathcal{X}}
\newcommand{\PL}[0]{\lambda}
\newcommand{\PLS}[0]{\lambda^{'}}
\newcommand{\PLSET}[0]{\Lambda}

\IEEEoverridecommandlockouts

\begin{document}
%
\title{Polygon Area Decomposition Using a Compactness Metric}





\author{\IEEEauthorblockN{Mariusz Wzorek\IEEEauthorrefmark{1}, Cyrille Berger\IEEEauthorrefmark{1}, Patrick Doherty\IEEEauthorrefmark{2}\IEEEauthorrefmark{1}}
\IEEEauthorblockA{\IEEEauthorrefmark{1} Department of Computer and Information Science, Link{\"o}ping University, Link{\"o}ping, Sweden}
\IEEEauthorblockA{\IEEEauthorrefmark{2} School of Intelligent Systems and Engineering, Jinan University (Zhuhai Campus), Zhuhai, China}
\thanks{This work has been supported by the ELLIIT Network Organization for Information and
Communication Technology, Sweden; the Swedish Foundation for Strategic
Research SSF (Smart Systems Project RIT15-0097); and the Wallenberg
AI, Autonomous Systems and Software Program: WASP WARA-PS
Project. The 3rd author is supported by a  RExperts Program Grant 2020A1313030098 from the Guangdong Department of Science and Technology, China.}
}

\maketitle

\thispagestyle{plain}
\pagestyle{plain}


\begin{abstract}

In this paper, we consider the problem of partitioning a polygon into a set of connected disjoint sub-polygons, each of which covers an area of a specific size. The work is motivated by terrain covering applications in robotics, where the goal is to find a set of efficient plans for a team of heterogeneous robots to cover a given area. Within this application, solving a polygon partitioning problem is an essential stepping stone. Unlike previous work, the problem formulation proposed in this paper also considers a compactness metric of the generated sub-polygons, in addition to the area size constraints. Maximizing the compactness of sub-polygons directly influences the optimality of any generated motion plans. Consequently, this increases the efficiency with which robotic tasks can be performed within each sub-region. The proposed problem representation is based on grid cell decomposition and a potential field model that allows for the use of standard optimization techniques.
A new algorithm, the AreaDecompose algorithm, is proposed to solve this problem. The algorithm includes a number of existing and new optimization techniques combined with two post-processing methods. The approach has been evaluated on a set of randomly generated polygons which are then divided using different criteria and the results have been compared with a state-of-the-art algorithm. Results show that the proposed algorithm can efficiently divide polygon regions maximizing compactness of the resulting partitions, where the sub-polygon regions are on average up to 73\% more compact in comparison to existing techniques. 

\end{abstract}


%
\IEEEpeerreviewmaketitle

\section{Introduction}\label{sec:intro}

The problem of partitioning a polygon into a set of non-overlapping primitive shapes whose union is equivalent to the original polygon has been extensively studied. Such problems arise in a vast number of applications. Examples include processes of Very Large Scale Integration (VLSI) circuit design~\cite{Asano:83}\cite{Asano:86}, pattern recognition~\cite{feng1975decomposition}, image processing~\cite{moitra1991finding}, database systems~\cite{lodi1978two}, or parallel computing~\cite{area-constr-christou1996optimal}, to name a few. 
The broad application spectrum of polygon partitioning results in a wide range of problem formulations where polygons are decomposed into sets of triangles, rectangles, and trapezoids, with or without additional size, perimeter, or other constraints. 

In recent work~\cite{Hert1998PolygonAD}, the \emph{area partitioning problem} has been formulated with a focus on applications to terrain covering in robotics.
In this case, the general goal of approaches designed to solve the area partitioning problem is to calculate a set of sub-regions that can be assigned to individual robots for tasks such as data collection, surveillance or exploration.
In order to provide efficient solutions for a team of heterogeneous robots, the sub-region sizes should be defined relative to the individual platform capabilities.
As a motivating example, we consider the problem of terrain covering using a fleet of Unmanned Aerial Vehicles (UAVs).
In this case, one can assume a number of UAVs collaborating to achieve a global collaborative goal such as gathering image sensor data from the given environment using their camera sensors. 
Each UAV would have its particular capabilities that would limit its range, speed and altitude in addition to constraints imposed by its sensor characteristics, such as field-of-view or image resolution.
In order to achieve the common goal of covering an entire environment efficiently, the whole region described as a polygon can be divided into set of sub-regions each of whose area size corresponds to a particular UAV's relative capabilities and sensor constraints.
\begin{figure}[!t]
	\centering
	\includegraphics[width=0.94\columnwidth]{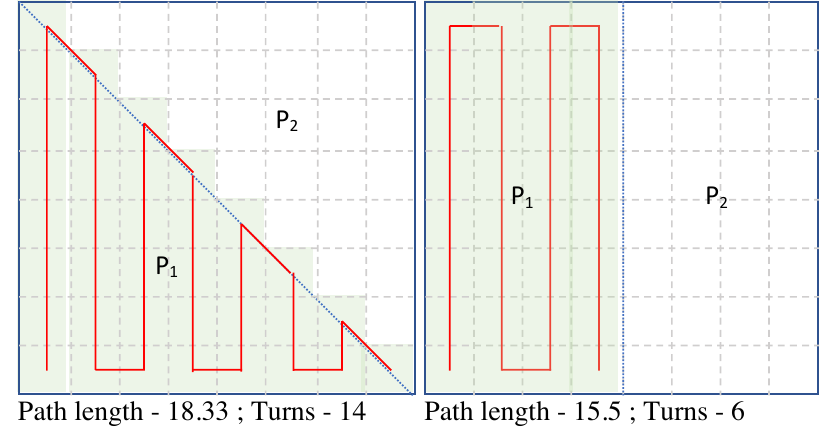}
	\caption{Terrain coverage examples.}
	\label{fig:terrain_coverage}
\end{figure} 
Fig.~\ref{fig:terrain_coverage} presents two examples where an environment described as a polygon is divided into two sub-regions of equal sizes ($P_1$ and $P_2$) to be assigned for a data gathering mission among two homogeneous UAVs. The partitioning of the polygon can be done arbitrarily as long as the area sizes are equal. In the examples provided, one division results in a triangular shape sub-region (left in the figure) and the other in a rectangular sub-region (right in the figure). The flight paths generated by a motion planner that cover these two different sub-regions will differ in efficiency as the path covering the triangular region is longer and requires more turns which will result in longer execution times and additional power or fuel usage. Additionally, in the triangular case the UAV covering region $P_1$ will unnecessarily overlap with region $P_2$. Ideally, one would want to minimize the overlap and provide regions that can be efficiently traversed.
In the general case, the shape of the resulting sub-region will influence these two factors as demonstrated in previous studies, for example in~\cite{Skorobogatov2021}. One of the measures that can be used to differentiate the properties of these shapes is the polygon \emph{compactness} or \emph{fatness} metrics.


Several compactness measures have been proposed in the literature, most notably with applications to problems related to Geographic Information Systems (GIS)~\cite{li2013}. One example is redistricting (i.e. drawing electoral district boundaries) for political elections in order to avoid the problem of gerrymandering (\cite{polsby:91,schwartzberg65,Reock:61}).
\begin{figure}[!t]
	\centering
	\includegraphics[width=0.94\columnwidth]{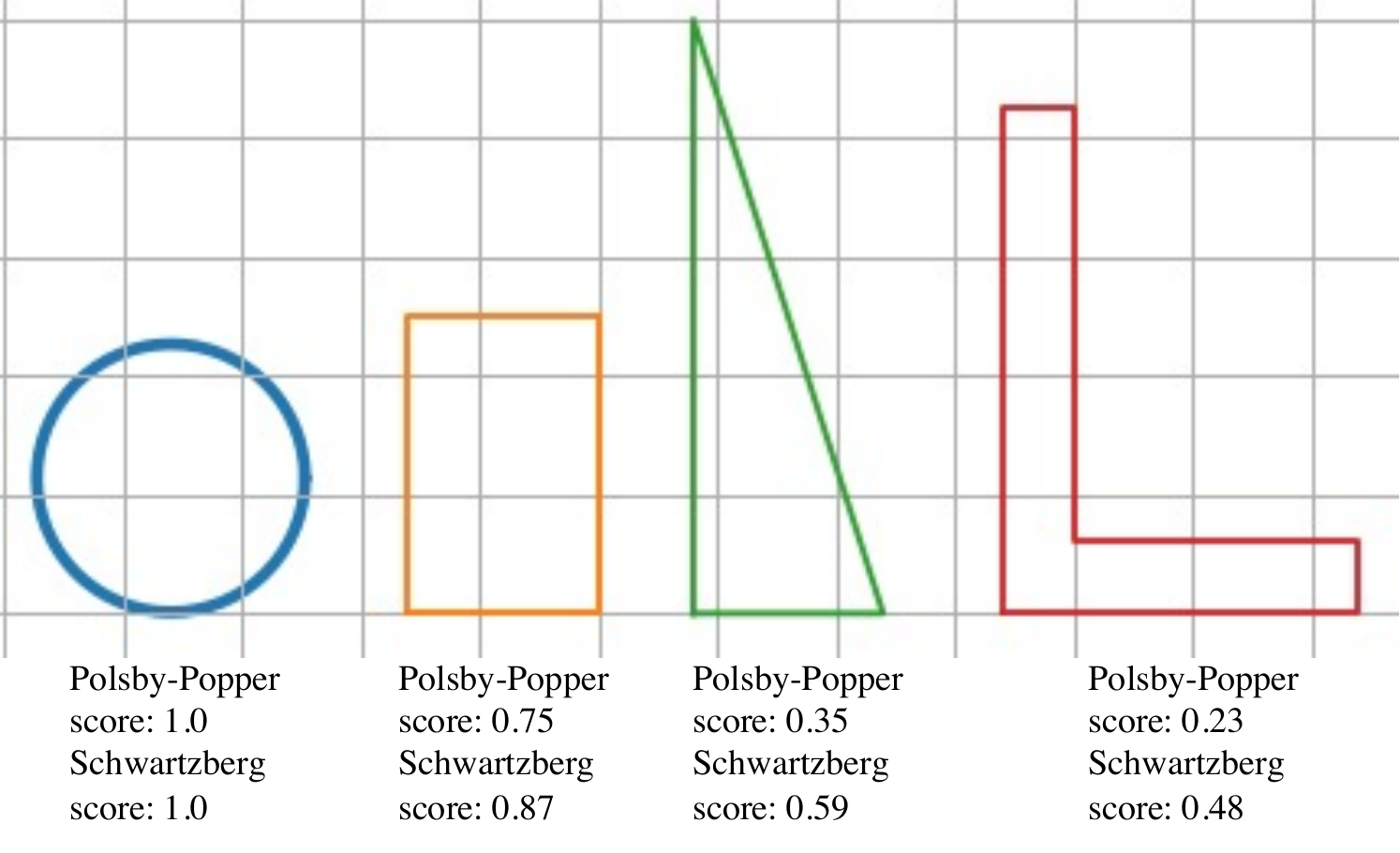}
	\caption{Examples of compactness scores for different shapes with equal areas.}
	\label{fig:compactness_scores_examples}
\end{figure} 
These measures often compare geometric properties of the polygon (e.g. area, perimeter) to the properties of a base geometric shape (e.g. circle) and take values from 0 to 1. Examples of two compactness scores, the Polsby-Popper~\cite{polsby:91} and Schwartzberg~\cite{schwartzberg65} scores for different shapes with equal area sizes are presented in Fig.~\ref{fig:compactness_scores_examples}. The base geometric shape used in these measures is a circle, which is considered the most compact shape. Thus the score for a circle is equal to 1.0 and as the shape compactness degrades the score values decrease.

In this paper we propose a novel area partitioning problem formulation where the goal is to divide a polygon into a set of disjoint connected sub-polygons with the focus on maximizing the sub-polygons' compactness as well as satisfying the area size constraints. The problem formulation is based on grid cell discretiztion and a potential field model.
Additionally, an algorithm is proposed for solving the new area partitioning problem efficiently. The algorithm includes a number of optimization techniques combined with two post-processing methods.


Although the terrain covering problem in UAV domains is used as a motivating application in this paper, the proposed approach is generic in nature. Other application domains where the shape characteristics are important could benefit from using the new area partitioning problem formulation and the proposed algorithms. Additionally, the proposed representation allows for applying the approach to GIS problems related to zoning or redistricting. An example of such an application is presented in~\Cref{seq:redistricting}.



\subsection{Related work}


One of the most common forms of polygon decomposition considered in the literature is polygon triangulation, where the goal is to decompose a polygon into a set of triangles under different settings. For example, in~\cite{triangles-de1992line} a Delaunay triangulation is used. Work presented in~\cite{triangles-baker1988nonobtuse,triangles-bern1995linear} considers decomposition using nonobtuse triangles. Other simple shapes have been used in the polygon decomposition problems as well. These include problems of polygon partitioning into trapezoids~\cite{Asano:83,Asano:86}, rectangles~\cite{levcopoulos1984bounds}, convex polygons~\cite{greene1983decompositionConvex,hertel1983fast,lien2006approximate} or star-shaped polygons~\cite{keil1985decomposing}. However, these algorithms  typically focus on finding a set of simple shapes that constitute a given polygon without considering additional constraints such as area sizes.





Polygon decomposition with area constraints where sub-polygons have equal area sizes have been proposed in~\cite{area-constr-page1992area,area-constr-christou1996optimal,Adjiashvili:10}. However, these approaches are limited to finding sub-polygons of equal area sizes thus restricting their use in the applications considered in this work.


A more general approach is presented in ~\cite{Hert1998PolygonAD}, where the area sizes can be defined arbitrarily. The authors, \emph{Hert\&Lumelsky}, propose a polynomial-time algorithm that utilizes a divide-and-conquer strategy combined with a sweep-line approach. The algorithm generates sub-polygons that are guaranteed to  satisfy the exact area size constraints. In comparison, the approach presented in this paper is based on a  representation of the area partitioning problem based on grid cell discretization. This allows for applying standard optimization techniques to find solutions to the problem that includes compactness metrics in addition to satisfying the area size constraints. Empirical comparisons between this algorithm and our approach are made in Section~\ref{sec:evaluation}.

Several approaches that include the area partitioning algorithms has been proposed in the application domain of terrain covering using robotic systems ~\cite{Skorobogatov2021},~\cite{berger:2016},~\cite{Agarwal:06},~\cite{agarwal2007rectilinear}. In~\cite{Skorobogatov2021},~\cite{berger:2016}, methods for generating flight plans for multiple UAVs are proposed based on algorithm in~\cite{Hert1998PolygonAD}. In~\cite{Skorobogatov2021} compactness of sub-polygons is measured and the results show that using compact sub-polygons yields more efficient area covering missions as plans generated for each sub-polygon assigned to a UAV exhibit a lower number of turns and shorter flight paths. In~\cite{berger:2016} motion plans generated for each UAV were optimized for flight time using a camera or laser range finder sensor model. In~\cite{Agarwal:06},~\cite{agarwal2007rectilinear} an approach for parallel region coverage using multiple UAVs is presented. The polygon decomposition method proposed is unfortunately limited to rectilinear polygons. Although the approaches presented in these approaches consider application of terrain covering using robotic systems, the underlying methods used for polygon decomposition are either based on the algorithm presented in~\cite{Hert1998PolygonAD} or are limited to rectilinear polygons and do not include compactness metrics as the optimization objective. As such, these approaches are not directly comparable to the work presented here with the exception of~\cite{Hert1998PolygonAD}, but rather the new specification of the area partitioning problem and the new algorithms presented can be incorporated in these solutions.

\subsection{Contributions}
The work described in this paper includes the following contributions:
\begin{itemize}
    \item A new area partitioning problem representation using a grid cell discretization and potential field model~\cite{pf-robotics-Khatib1990} that allows for applying standard optimization techniques. Solutions acquired by using the new problem formulation trade-off accuracy of satisfying area size constraints while allowing for maximization of sub-polygon compactness.
    \item A new algorithm for solving the area partitioning problem which includes a number of optimization algorithms combined with two post-processing methods. Integrated algorithms include a new heuristic-based method (the Potential Field Heuristic), the Covariance Matrix Adaptation Evolution Strategy (CMA-ES)~\cite{Hansen2006}, and Random Search optimization techniques.
    \item An empirical evaluation of the proposed algorithm on a set of randomly generated non-convex polygons and comparison with an existing state-of-the-art technique~\cite{Hert1998PolygonAD}.
\end{itemize}
The structure of this paper is as follows.
First, a problem definition is provided in \Cref{sec:problem}.
\Cref{sec:algorithm} describes the proposed polygon area decomposition algorithm.
The results of experimental evaluation are presented in \Cref{sec:evaluation}.
Finally, the conclusions with a short description of future work are presented in \Cref{sec:conclusions}.


\section{Problem definition}\label{sec:problem}

Formally, the problem considered in this paper can be described as follows. Given a simple\footnote{a polygon with no self-intersection} polygon $\mathcal{P}$, decompose the polygon into $n$ disjoint sub-polygons $\mathcal{P}_1$ to $\mathcal{P}_{n}$. The area of each sub-polygon ($\mathcal{A}(\mathcal{P}_i) : i\in I_n \in \{1,\ldots,n\}$) is specified by a weight ($\omega_i \in \Omega$). Weights define area sizes as a proportion of the total area of the polygon $\mathcal{A}(\MP)$. Thus, the area partitioning problem is defined as follows:
\begin{alignat}{3}
    & \mathcal{P}                         && = \bigcup\limits_i \mathcal{P}_i          && \\
    & \mathcal{P}_i \cap \mathcal{P}_j    && = \emptyset                               && \quad \forall i \neq j \\
    & \mathcal{A}(\mathcal{P}_i)          && = \omega_i \cdot \mathcal{A}(\mathcal{P}) && \quad \forall i  \\
    & \sum_i \omega_i                     && = 1                                       &&
\end{alignat}
Additionally, to take shape characteristics into account, the generated sub-polygons $\MP_i$ should maximize the compactness score of each sub-polygon. 
Since sub-polygons are represented as sets of polylines each defined by its vertices (points) the state space of the problem that needs to be considered is infinite. In the approach presented in this paper, the problem representation is relaxed by discretizing the polygon $\mathcal{P}$ into a finite size set of grid cells. Using such representation allows for redefining of the infinite domain problem into a finite domain constraint optimization problem. 

Let $\MV$ be a finite set of grid cells $\nu_k$ of size $s$ that cover the polygon $\MP$. A grid cell $\nu_k$ is defined as a tuple $\left<c^\nu_k, \alpha^\nu_k \right>$ where $c^\nu_k =( x^c_k, y^c_k) \in \mathbb{R}^2$ is the center of the grid cell, and $\alpha^\nu_k$ is its contributing area. The area is calculated as the sum of the intersections of the grid cells with the polygon $\mathcal{P}$. Thus, the area values for the grid cells not fully included in the polygon (cells close to polygon edges) will be bounded by $\alpha^\nu_k \leq s^2 $. Formally, this is defined as:
\begin{align}
    \MV &= \{\nu_k : k \in I_k = \{1,\ldots,m\}\} \\
    \nu_k &= \left<c_k^\nu,\alpha^\nu_k\right> \\
    \alpha^\nu_k &= \MA(\nu_k \cap \MP) \\
    \MP &= \bigcup \nu_k \\
    \MA(\MP) &= \sum_{} \alpha_k^\nu
\end{align}

\begin{figure}[!t]
    \centering
    \includegraphics[width=1.0\columnwidth]{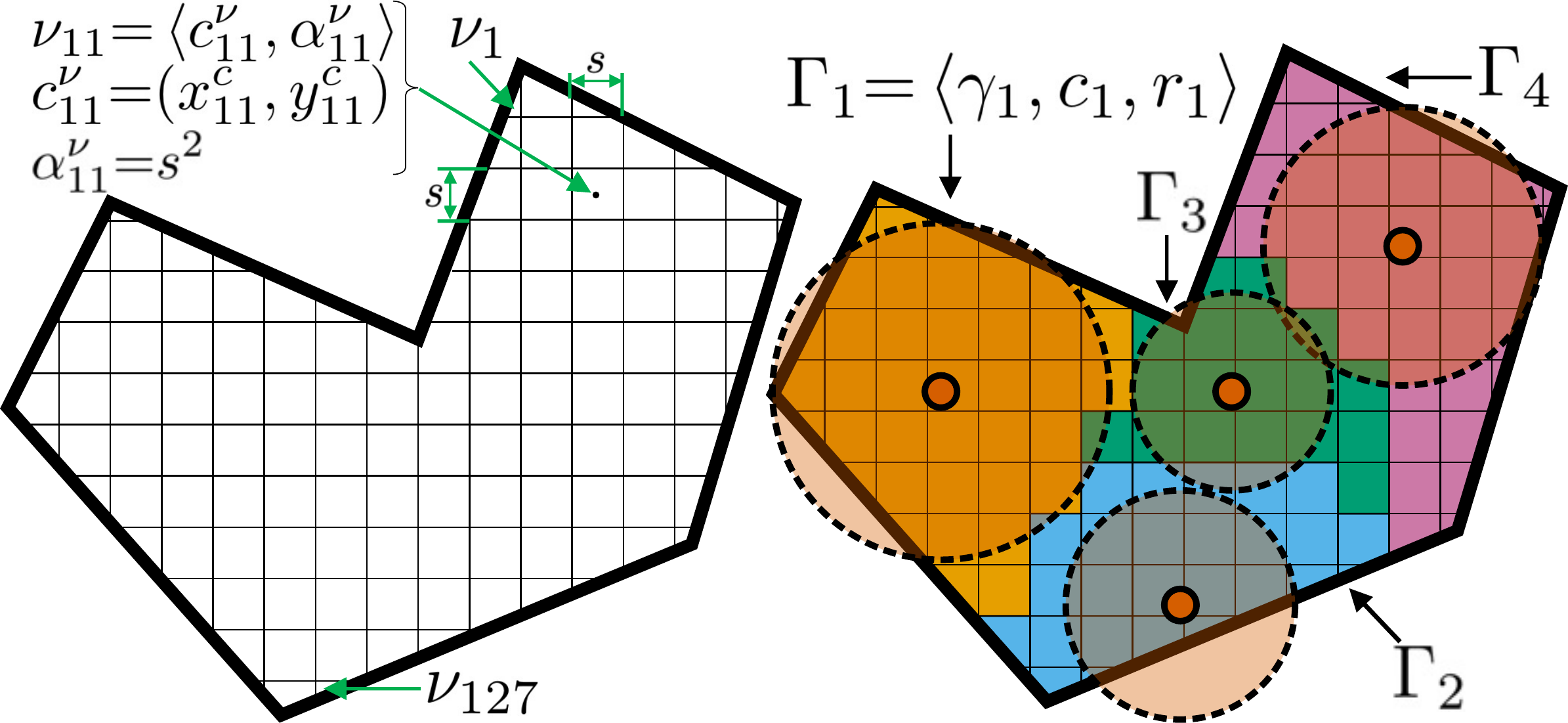}
    \caption{Illustrative example of a discretized area partitioning problem definition. Division of polygon $\MP$ into a finite set of grid cells $\MV$ (left). Potential fields associated with each sub-polygon $\MP_i$ represented as partitions $\Gamma_i$ (right).}
    \label{fig:grid_division}
\end{figure}


An illustrative example of grid cell decomposition is presented in Fig.~\ref{fig:grid_division} (left), where the original polygon has been decomposed into a set of 127 grid cells ($\MV=\{\nu_1,\ldots,\nu_{127}\}$).
The original problem of area polygon partitioning can be represented as finding $n$ disjoint subsets of $\MV$ subject to area constraints.
A potential field model is used to compute an assignment of cells to each subset of $\MV$, where each sub-polygon is represented as an attractive potential defined by a circle with its center position $c_i=(x_i,y_i)$ and a radius $r_i$. The radius corresponds to the attractive force value of the potential.
Let $\Gamma_i \in \Gamma$ : $i\in I_n \in \{1,\ldots,n\}$ represent a sub-polygon partition.
$\Gamma_i$ is defined as a tuple $\left<\gamma_i, c_i, r_i\right>$, where $\gamma_i \subset \MV$ is a set of grid cells that belong to the partition, and $c_i$ and $r_i$ define the attractive force associated with it.
An example of four sub-polygon partitions for a simple polygon is presented in Fig.~\ref{fig:grid_division} (right, using color-coding for each partition).

Allocation of grid cells to a particular sub-polygon partition $\Gamma_i$ is defined by the following function:
\begin{align}
 a(k) = \argmax\limits_i  \dfrac{d(c_i, c^\nu_k)}{r_i}\label{eq:assignArgmax}
\end{align}
where parameter $k$ is a grid cell index, $d(c_i,c^\nu_k)$ is the euclidean distance between the potential field center $c_i$ and a grid cell center $c^\nu_k$, and $r_i$ is the potential field radius. Thus, using such representation a sub-polygon $\mathcal{P}_i$ is defined as a union of grid cells that has been assigned to the polygon partition:
\begin{align}
    \mathcal{P}_i = \bigcup_{\nu_j \in \gamma_i} \nu_j \label{eq:partitionsToPoly}\\
    \gamma_i = \{\nu_j : a(j)=i \}\label{eq:assignToPartition}
\end{align}

Given the discretized representation of the original area partitioning problem, the following bi-objective function is used to define it as an optimization problem that captures both the area size constraints and shape characteristics:

\begin{align}
    f \left( \cal{X} \right) & = f_{area}\left( \cal{X} \right) - f_{shape}\left( \cal{X} \right) \label{eq:objective} \\
    f_{area}\left( \cal{X} \right)^2 & = \frac{1}{n}\sum_i^n  \mathcal{A}^{err}(\mathcal{P}_i, \omega_i, \mathcal{P})^2\label{eq:objective1} \\
    f_{shape}\left( \cal{X} \right) & = \frac{1}{n}\sum_i^n  f_c\left(\MP_i\right)\label{eq:objective2}
\end{align}
\noindent
where $f:R^n\mapsto  \mathbb{R}$ 
(each $x \in \cal{X}$ is a tuple $\left\langle c_i, r_i\right\rangle $), and $f_c(\MP_i) \in [0,1]$ is a compactness score for sub-polygon $\MP_i$.
The $\mathcal{A}^{err}(\mathcal{P}_i, \omega_i, \mathcal{P})$ is a function of area size error for each sub-polygon defined as:
\begin{equation}
  \mathcal{A}^{err}(\mathcal{P}_i, \omega_i, \mathcal{P}) = 
    \frac{\mathcal{A}\left(\mathcal{P}_i\right) - \omega_i \cdot \mathcal{A}\left(\mathcal{P}\right)}{\omega_i \cdot \mathcal{A}\left(\mathcal{P}\right)}
\end{equation}
Minimizing the objective function $f_{area}\left( \cal{X} \right)$ ensures that each sub-polygon meets its target area size. Maximizing the second objective function $f_{shape}\left( \cal{X} \right)$ ensures that sub-polygon compactness is maximized.
Intuitively, when $f(\MX)=-1$, the set of sub-polygons meet the exact area size requirements defined by weights in $\Omega$ and each sub-polygon compactness score is equal to 1.

The optimization problem for solving the area partitioning problem using the discretized representation is defined as:
\begin{subequations}
	\label{eq:optim}
	\begin{flalign}
	\underset{\mathcal{X}}{\text{min}}
	& \quad f(\mathcal{X}) \label{eq:cost}\\
	        \text{s.t.} 
	& \quad \mathcal{P} = \bigcup\limits_i \mathcal{P}_i \label{eq:const1}\\
	& \quad \mathcal{P}_i \cap \mathcal{P}_j = \emptyset \quad \forall i \neq j \label{eq:const2}\\
	& \quad \left|\mathcal{A}^{err}(\mathcal{P}_i, \omega_i, \mathcal{P})\right| \leq \tau \quad\forall i \label{eq:const3}\\
   & \quad \sum_i \omega_i = 1 \label{eq:const4}
	\end{flalign}
\end{subequations}
\noindent
where a sub-partition $\MP_i$ is the result of the cell assignment of $\mathcal{X}$ according to \cref{eq:partitionsToPoly}.
Constraints defined in the optimization problem ensure that: (1) the sub-polygons are disjoint (\cref{eq:const2}); (2) the sub-polygons cover the whole area defined by the original polygon $\MP$ (\cref{eq:const1}); and (3) the weights defined for area sizes are a proportion of the total area of polygon $\MP$ (\cref{eq:const4}).
Satisfying the area size constraints for each sub-partition is not exact since the proposed representation uses a set of grid cells of finite size. Let $\tau$ be defined as an upper bound on the area size error for each partition. The constraint that enforces the bounds on area size errors is defined by~\cref{eq:const3}.

In order to apply general optimization methods for solving the polygon decomposition problem defined above, the objective function including penalties for constraint violations is defined as follows:

\begin{align}
\underset{\mathcal{X}}{\text{min}} (f(\mathcal{X}) + \pi(\mathcal{X})); \;\;  \pi(\mathcal{X})=(\pi_c \cdot \pi_p(\mathcal{X}))^2 \label{eq:opt_with_penalties}\\
\pi_p(\mathcal{X}) = \max\left(0, \max\limits_i\left( \left| \mathcal{A}^{err}(\mathcal{P}_i, \omega_i, \mathcal{P}) \right| \right) - \tau\right)
\end{align}

\noindent where $\pi_c$ is a penalty coefficient chosen empirically.

\begin{figure*}[!th]
    \centering
    \includegraphics[width=0.8\textwidth]{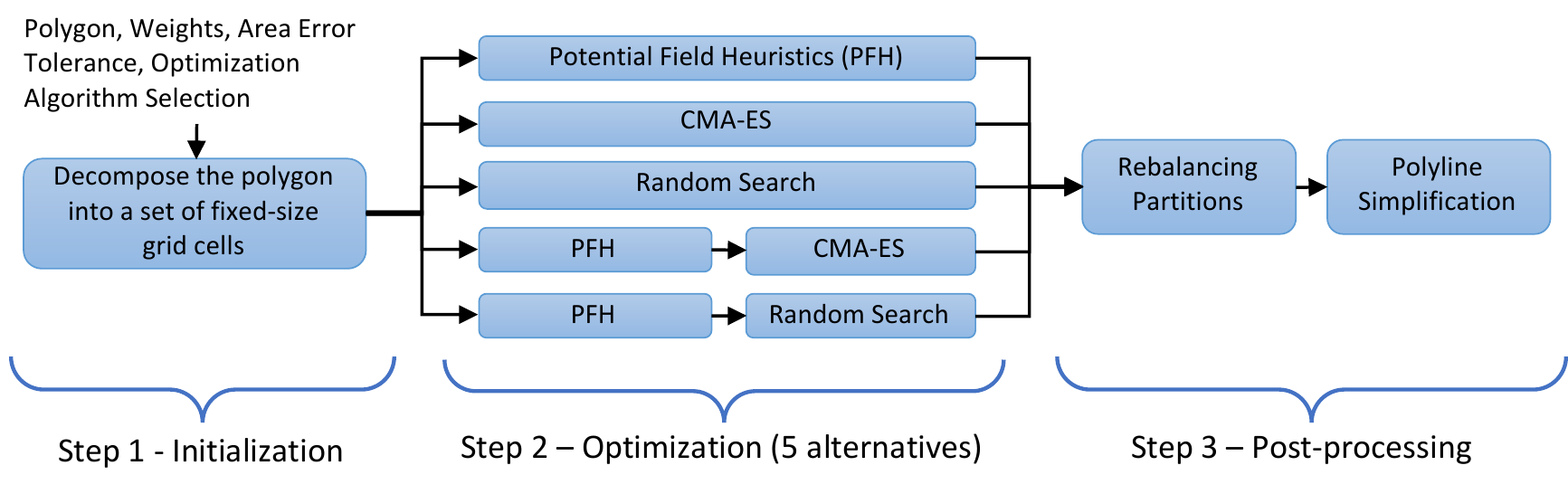}
    \caption{The AreaDecompose algorithm overview.}
    \label{fig:algorithm_overview}
\end{figure*}
\section{Polygon Area Decomposition Algorithm}\label{sec:algorithm}

\begin{figure*}[!th]
    \centering
    \subfloat[Reassignment of a cell that causes non-continuous partition boundary. Partitions before and after the reassignment are shown on the left and right side, respectively.]{
        \includegraphics[width=0.45\textwidth]{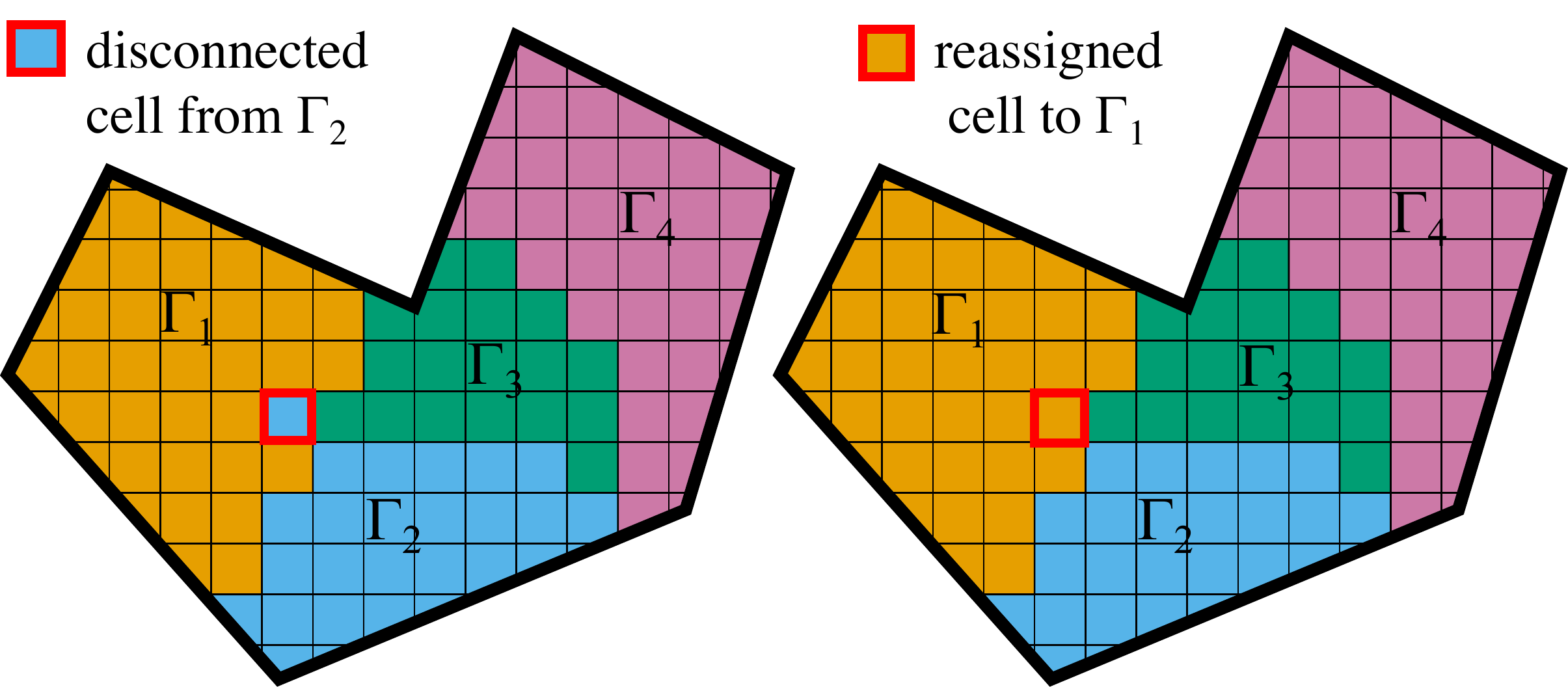}
        \label{fig:rebalancing_partitions_1}
    }\quad
    \subfloat[Reassignment of cells to minimize area size errors. Candidate cells from partition $\Gamma_1$ which has been assigned too many cells, considered for reassignment to partition $\Gamma_2$ are shown on the left side. Result of the reassignment is shown on the right side. ]{
        \includegraphics[width=0.45\textwidth]{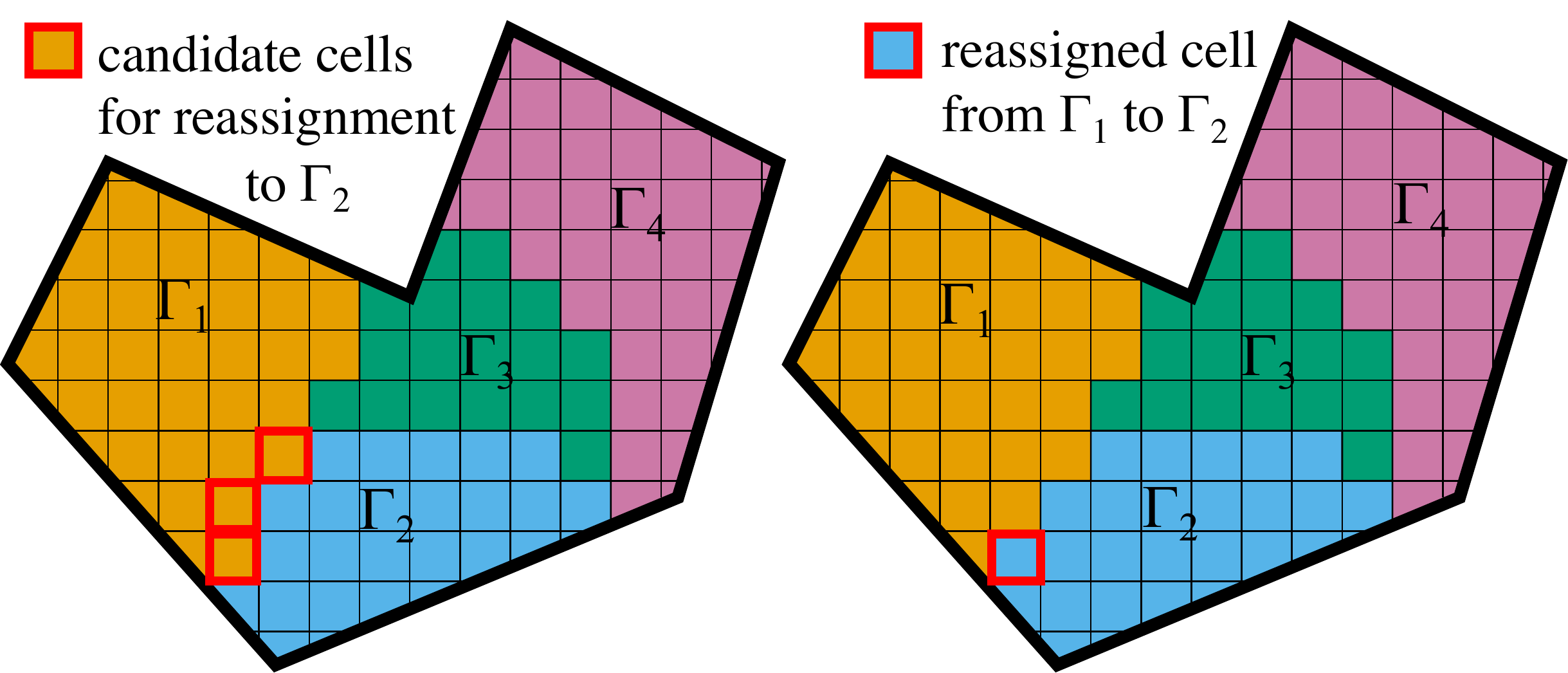}
        \label{fig:rebalancing_partitions_2}
    }
    \caption{An illustrative example of applying the rebalancing partitions post-processing steps.}
\end{figure*}

In this section, a description of the proposed algorithmic framework is presented. Fig.~\ref{fig:algorithm_overview} provides an overview of the \emph{AreaDecompose} algorithm.
Given a polygon description and a set of weights, the algorithm starts by dividing the polygon into a set of grid cells with a fixed size (Step 1). In Step 2, one of the five alternative methods for optimization is applied to assign grid cells to each sub-polygon. The first alternative, a \emph{Potential Field Heuristics} (PFH) method, uses a heuristic-based approach for finding optimal values of attractive force fields for each sub-polygon. The second and third alternatives apply generic optimization techniques, namely the \emph{Covariance Matrix Adaptation Evolution Strategy} (CMA-ES) and \emph{Random Search} (RS). The final two alternatives in the optimization step combine the heuristic-based PFH algorithm with the generic CMA-ES and RS approaches. In this case, the PFH provides an initial solution that is then refined by applying the CMA-ES or RS optimization techniques.
The final step of the \emph{AreaDecompose} algorithm (Step 3) includes two post-processing procedures.
The first procedure deals with the problem of removing potential disconnected cells causing invalid sub-polygon borders (Section~\ref{sec:rebalancing_partitions}, Fig.~\ref{fig:rebalancing_partitions_1}) in the generated solutions. Additionally, a heuristic-base procedure is applied that tries to improve values of area size errors for each partition. Sub-polygons are rebalanced by exchanging grid cells when possible to meet the area size constraints (Section~\ref{sec:rebalancing_partitions}, Fig.~\ref{fig:rebalancing_partitions_2}). The second post-processing step applies a polyline simplification algorithm to smooth out sub-polygon borders that define the final sub-polygons (Section~\ref{sec:polyline_simplification}, Fig.~\ref{fig:polyline_simplification_overview}).


\begin{algorithm}
	\DontPrintSemicolon
	\KwIn{$\mathcal{P},\Omega$}
	\KwOut{$\left<\mathcal{P}_1, \ldots, \mathcal{P}_n\right> $}
	\Parameter{$\tau$, \emph{optAlg}}
	$s = \sqrt{\tau \cdot min(\Omega) \cdot \mathcal{A}(P)}$\;
	$\mathcal{V} \gets$ \emph{divideIntoGrids}$(\mathcal{P},s)$\;
	\For{$\Gamma_i$ \textup{\textbf{in}} $\Gamma$} {
	    $\gamma_i \gets \emptyset$ \;
	    $c_i \gets$ \emph{initOnBoundary}$(\mathcal{P})$ \;
	    $r_i \gets$ \emph{expectedRadius}$(\omega_i,\mathcal{P})$\Comment*[r]{\cref{eq:initial_ri_radius}}
	}
	\If {\{"PFH"\} $\cap$ \textit{optAlg}$ \neq \emptyset$}{
	    $\Gamma \gets$ \emph{PFH}$(\mathcal{P},\Omega,\mathcal{V},\Gamma)$
	    }
	\If {\{"CMA-ES", "RandomSearch"\} $\cap$ \textit{optAlg}$ \neq \emptyset$}{
	    $\Gamma \gets$ \emph{runOptimization}$($\emph{optAlg}$,\mathcal{P},\Omega,\mathcal{V},\Gamma)$
	}
	$\Gamma \gets$ \emph{rebalancePartitions}$(\Gamma)$\;
	$\left<\mathcal{P}_1, \ldots, \mathcal{P}_n\right> \gets$  \emph{partitionsToPolygon}$(\Gamma)$\Comment*[r]{\cref{eq:partitionsToPoly}}
	$\left<\mathcal{P}_1, \ldots, \mathcal{P}_n\right> \gets$ \emph{simplifyBorders}$(\MP,\left<\mathcal{P}_1, \ldots, \mathcal{P}_n\right>, s)$\;
	
	\Return{$\left<\mathcal{P}_1, \ldots, \mathcal{P}_n\right>$}\;
	\caption{{\sc AreaDecompose}}
	\label{algo:AreaDecompose}
\end{algorithm}

The \emph{AreaDecompose} algorithm is presented in \Cref{algo:AreaDecompose}.
An input to the algorithm is the polygon $\MP$ and a set of weights $\Omega$ defining the relative area sizes of generated sub-polygons as proportions of the total area of $\MP$. The number of sub-polygons $n$ that the algorithm will calculate is equal to $|\Omega|$.
Other parameters used during the calculations are the minimum tolerance $\tau$ value for an error on area size of each sub-polygon, and a set of flags $optAlg$ defining which optimization techniques to apply.

The algorithm starts by decomposing the polygon $\MP$ into a set of grid cells $\mathcal{V}$ (see Fig.\ref{fig:grid_division}, Lines 1-2). The grid cell size $s$ used for decomposition is calculated based on the requested upper bound on the area size error $\tau$ and the smallest partition size defined by weights in $\Omega$ as follows:
\begin{align}
    s = \sqrt{\tau \cdot min(\Omega) \cdot \mathcal{A}(P)} \label{eq:grid_cell_size}
\end{align}

Such grid cell size selection criteria ensures that the area error requirements $\tau$ can be met.

Next, the $\Gamma_i$ partitions are initialized. The list of grid cells $\gamma_i$ assigned to partition $i$ is initialized as empty (Line 4). Initial centers of attractive potential fields $c_i$ for each partition is calculated as equidistant points on the perimeter of polygon $\MP$ (Line 5). The radius of each potential field circle $r_i$ is initialized so that the area of the circle is equal to the expected area of the sub-polygon defined by its weight $\omega_i$:

\begin{align}
    r_i \leftarrow \sqrt{\dfrac{\omega_i \cdot \mathcal{A}(\mathcal{P})}{\pi}} \label{eq:initial_ri_radius}
\end{align}

Depending on the algorithm configuration flags ($optAlg$) the algorithm executes a combination of methods that include a heuristic-based algorithm PFH, or optimization methods based on CMA-ES or RandomSearch (Lines 7-10). These are the main methods that search for $\langle c_i, r_i\rangle$ potential field configurations that yield minimal area errors and maximize compactness of each partition as defined by the optimization problem in \cref{eq:cost}-\cref{eq:const4}.

After the main part of the algorithm is executed, the resulting configurations of sub-polygons $\Gamma$ are further optimized by a post-processing step that tries to improve cell assignments (Line 11).
The \emph{rebalancePartitions} function considers two factors. On rare occasions the disjoint sets of grid cells generated by the algorithms can result in non-continuous partition borders (Fig.~\ref{fig:rebalancing_partitions_1}). Additionally, the assignment of cells is checked if further improvement of area size errors can be made by switching grid cells between neighboring partitions (Fig.~\ref{fig:rebalancing_partitions_2}).

Next, the resulting partitions defined by $\gamma_i$ are transformed into a set of polylines describing each sub-polygon $\MP_i$ (Line 12).
The final step of the algorithm (Line 13) applies the \emph{simplifyBorders} function that simplifies polylines resulting from a grid cell decomposition (zigzag) line segments.

Details describing the main optimization and post-processing methods are presented in the remainder of this section. 
\begin{algorithm}
	\DontPrintSemicolon
	\KwIn{$\mathcal{P},\Omega,\mathcal{V},\Gamma$}
	\KwOut{$\Gamma$}
	\Parameter{$\tau$, \emph{MaxIter}}
	\emph{Finished} $\gets false$\;
	\emph{Iter} $\gets 1$\;
	\While{$\neg$Finished} {
		$\xi \gets \frac{MaxIter - Iter}{2 \cdot MaxIter}$\;
		$\Gamma \gets$ \emph{updateRadiuses}$(\mathcal{P},\omega_i,\mathcal{P}_i,\xi,\Gamma)$\Comment*[r]{eq.(\ref{eq:pfh_updateRadius1}-\ref{eq:pfh_updateRadius3})}
		\For{$v_k$ \textup{\textbf{in}} $\mathcal{V}$} {
		    $j \gets$ \emph{assignCells}$(\Gamma, \mathcal{V})$\Comment*[r]{\cref{eq:assignArgmax}}
		    $\gamma_j = \gamma_j \cup {v_k}$\Comment*[r]{\cref{eq:assignToPartition}} 
		}
		$\Gamma \gets$ \emph{calculateCenters}$(\Gamma)$\Comment*[r]{\cref{eq:pfhCentroid}}
	    
	    \emph{Iter} $\gets$ \emph{Iter}$+1$\;
	    \If { \textit{Iter} $==$ \textit{MaxIter} \textup{\textbf{or}} $\forall \Gamma_i \in \Gamma: |\mathcal{A}(\Gamma_i)-\omega_i \cdot \mathcal{A}(\mathcal{P})| \leq \tau $}
		{
		\emph{Finished} $\gets true$
		}
	}
	\Return{$\Gamma$}\;
	\caption{{\sc PFH}}
	\label{algo:PFH}
\end{algorithm}

\subsection{Potential Field Heuristics}\label{sec:pfh}
The Potential Field Heuristics (PFH) is an iterative method that searches for the optimization parameters $c_i$, $r_i$ associated with each sub-polygon partition using a simple heuristics. The algorithm is presented in \Cref{algo:PFH}. The constant parameters used by the algorithm are the lower bound of the area size errors $\tau$ for sub-polygons and the maximum number of iterations $MaxIter$. 


The general idea behind the PFH algorithm is to use a heuristic function to update radiuses of potential field circles ($r_i$) associated with sub-polygon partitions ($\Gamma_i$). These updates directly influence the grid cell assignments for each partition ($\gamma_i$,~\cref{eq:assignArgmax}), thus changing partition configurations. Centers of potential field circles ($c_i$) are calculated as partition centroids based on the updated grid cell assignments ($\gamma_i$).

The execution of the algorithm starts with the initialization of two auxiliary variables (Lines 1-2). The first one is a boolean flag ($Finished$) used to determine if a termination condition of the algorithm has been satisfied. The second variable is a simple iteration counter used in the main loop of the algorithm which is executed in Lines 3-12. Conceptually the main loop consists of four steps. The first step is the heuristic-based update of $r_i$ radiuses (Lines 4-5). The second step, includes updates of sub-polygon partitions ($\gamma_i$, Lines 6-8). In the next step centers of potential field circles ($c_i$) are calculated (Line 9). Finally, two termination conditions are evaluated (Lines 11-12). The first condition checks whether area sizes of the resulting sub-polygon partitions $\Gamma_i$ are within the specified error $\tau$. The second condition checks if the algorithm reached its maximum number of iterations $MaxIter$.

The heuristics used for updating the $r_i$ radius associated with each potential field circle executed by the $updateRadiuses$ function is defined by following equations:
\begin{align}
    \delta &= \xi \cdot \left( \dfrac{\mathcal{A}(\mathcal{P}_i)}{\omega_i \cdot \mathcal{A}(\mathcal{P})} - 1 \right)\label{eq:pfh_updateRadius1} \\
    \xi &= \frac{MaxIter - Iter}{2 \cdot MaxIter}\label{eq:pfh_updateRadius2}\\
    r_i &\leftarrow \dfrac{r_i}{\delta + 1}\label{eq:pfh_updateRadius3}
\end{align}
where $\xi$ is a linear dampening factor dependent on the maximum iteration parameter $MaxIter$ and the iteration counter $Iter$ of the algorithm. 
Thus, in the initial iterations of the execution, updates to the radius $r_i$ are more radical taking bigger step values. While towards the end of the algorithm execution the updates become more refined.

Centers of the attractive potential field force $c_i$ for each partition are calculated as partition centroids:
\begin{align}
    c_i = \dfrac{1}{|\gamma_i|}\sum_{\nu_j \in \gamma_i} c^\nu_j\label{eq:pfhCentroid}
\end{align}


\subsection{CMA-ES Optimization}\label{sec:CMAES}
The Covariance Matrix Adaptation Evolution Strategy (CMA-ES)~\cite{Hansen2006} is an evolutionary optimization technique that has gained popularity in recent years. It has been successfully applied to many problems where the gradient of the objective function cannot be computed directly and where numerical computation is expensive. Additionally, the CMA-ES includes methods to mitigate local minima problems. 
The strategy used in the algorithm can estimate the correlation between the optimization parameters and can detect when the parameters are independent.
The CMA-ES searches for parameters $x$ among a set of possible parameters $\MX$ minimizing the value of the objective function $f$ without any specific knowledge or restrictions on the function itself. For example, $f$ can be non-convex, multi-modal, non-smooth, discontinuous, noisy, or ill-conditioned.

For the area partitioning problem, the CMA-ES is used to minimize the bi-objective function defined in \cref{eq:opt_with_penalties},
where $f:R^{2n}\mapsto  \mathbb{R}$ 
(each $x \in \cal{X}$ is a tuple $\left\langle c_i, r_i\right\rangle $). The algorithm searches for a set of potential field centers and their radiuses that minimizes the sub-polygon area size errors and maximizes the sub-polygon compactness. 
During each optimization step, the CMA-ES selects a set of parameters $\mathcal{X}$ which defines the grid cell assignment (\cref{eq:assignArgmax}, \cref{eq:assignToPartition}) used to evaluate the objective function $f(\MX)$.

\subsection{Random Search Optimization}\label{sec:RS}
A simple alternative to the CMA-ES optimization proposed in this paper uses a stochastic random search approach. In this case the optimization parameters and the objective function are the same as in the case of the CMA-ES. The algorithm generates a number of random configurations of $\MX$, which are evaluated and the best is returned.

%

\subsection{Rebalancing Partitions}\label{sec:rebalancing_partitions}


The sub-polygon partitions $\Gamma_i$ generated by the main optimization methods are further improved by applying the \emph{RebalancePartitions} post-processing step which includes two procedures. 



The focus of the first procedure is to deal with non-continuous partition boundaries that may result from grid cell assignment.
A sub-polygon partition boundary is defined by vertices shared with the original polygon $\MP$ and edges of its grid cells that are neighbors to other sub-polygon partitions.
Sporadically the assignment of grid cells to sub-polygon partitions $\Gamma_i$ based on \cref{eq:assignArgmax} may create divisions that result in non-continuous partition boundaries. 
Transforming such boundaries directly into polyline representation used by final sub-polygons $\MP_i$ will create self-intersecting polygons.
A concept of \emph{disconnected} grid cells can be used to identify non-continuous partition boundaries.
A grid cell $\nu_k$ assigned to a partition $\Gamma_i$ is considered to be \emph{disconnected} if all of its neighboring cells (top, bottom, left, right) are assigned to other partitions than the partition in question.
\begin{align}
disconnected(\nu_k, \gamma_i) \Longleftrightarrow \forall \nu_j \in Neighbor(\nu_k)(\nu_j \notin \gamma_i)
\end{align}
\noindent where $\nu_j = \{neighbors(\nu_k)\}$ are top, bottom, left and right neighboring cells of $\nu_k$. An example of a disconnected cell is presented in Fig.~\ref{fig:rebalancing_partitions_1} (left), where one cell assigned to $\Gamma_2$ has only cell neighbors that belong to $\Gamma_1$ and $\Gamma_3$ partitions.

The procedure used to remedy the problem of non-continuous partition boundaries is defined as follows. For each \emph{disconnected} grid cell select a new sub-polygon partition until all grid cells are connected to their respective partitions. The selection of a new partition is based on the original assignment equation (\cref{eq:assignArgmax}) where the second best partition is consider that is also a neighbor of the cell in question.
Fig.~\ref{fig:rebalancing_partitions_1} shows a simple example of the reassignment. In this case, the disconnected cell can be assigned to either partition $\Gamma_1$ or $\Gamma_2$ as both of the partitions are neighbors of the cell. After applying the assignment equation (\cref{eq:assignArgmax}) the $\Gamma_1$ sub-polygon partition is selected.


The second process included in the \emph{RebalancePartitions} post-processing step uses simple heuristics to improve area size errors for sub-polygon partitions. For example, suppose a partition has received too many grid cells during the optimization. In that case, the redundant cells can potentially be reassigned to a neighbor partition with an insufficient number of grid cells. Thus the area size errors for both partitions are minimized.



In order to ascertain if a partition $\Gamma_i$ has been assigned too many grid cells or an insufficient amount of cells, an approximate expected number of cells ($N_c^i$) is calculated according to the following equation:
\begin{align}
 N_c^i &= round\left( \frac{\omega_i \cdot \MA(\MP)}{\MA(\nu_i)}\right)\label{eq:rebalancing_expected_cells} \\
 N_r^i  &= |\gamma_i| - N_c^i\label{eq:rebalancing_missing_or_redundant_cells}
\end{align}
\noindent where $N_r^i$ corresponds to the number of cells that are redundant or absent in the partition $\Gamma_i$.

The post-processing procedure starts with selecting a partition $\Gamma_a \in \Gamma$ with the largest area that has been assigned a redundant number of cells (i.e. $N_r^a>0$). All neighbor partitions of $\Gamma_a$ are evaluated according to \cref{eq:rebalancing_missing_or_redundant_cells} and the partition which is missing the most cells is selected ($\Gamma_b$) to receive a grid cell from $\Gamma_a$. The process of selecting the recipient partition $\Gamma_b$ is defined by the following equation:

\begin{align}
\Gamma_b = \argmin\limits_{\Gamma_j \in neighbors(\Gamma_a)} \left(N_r^j \right)
\end{align}


The candidate cell selected for reassignment is the cell that is the furthest from the $\Gamma_a$ partition center $c_a$.
The procedure is repeated until no further cell reassignments can be made.

An illustrative example of the procedure is shown in Fig.~\ref{fig:rebalancing_partitions_2}. Partition $\Gamma_1$ has the largest area and has been assigned too many cells while its neighbor partition $\Gamma_2$ has an insufficient number of grid cells. The procedure reassigns one grid cell from partition $\Gamma_1$ to $\Gamma_2$.



\subsection{Polyline Simplification}\label{sec:polyline_simplification}

\begin{figure}[!t]
\centering
\includegraphics[width=1.0\columnwidth]{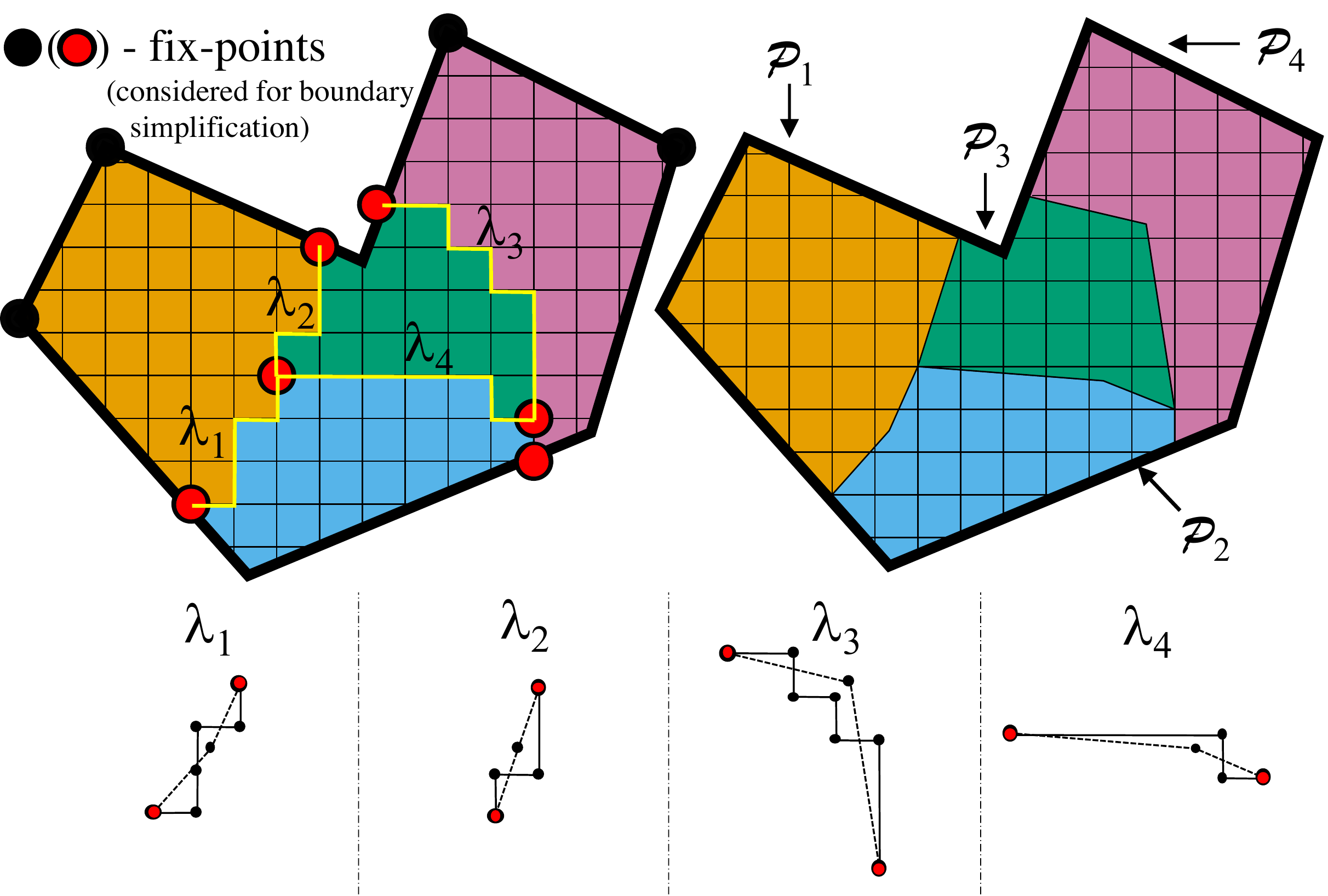}
\caption{An example of polyline simplification.}
\label{fig:polyline_simplification_overview}
\end{figure}

The borders between the sub-polygons $\Gamma_i$ generated by the grid cell assignment follows the edges of the cells. Because of that, the polylines defining the borders are in shape of a zigzag pattern and consists of an excessive number of segments.
The final post-processing step, \emph{simplifyBorders}, used in the \emph{AreaDecompose} algorithm, aims at simplifying these borders with \emph{k-point polylines} where the value of $k$ is minimal.
Fig.~\ref{fig:polyline_simplification_overview} depicts an example set of sub-polygons with their borders resulting from the grid cells assignment (top left) and their final borders after the simplification (top right). In this example, four borders have been simplified with 3-point polylines depicted at the bottom of the figure.

The \emph{simplifyBorders} procedure first identifies borders that need to be simplified. The grid corners shared between at least three partitions are considered to be fix-points together with any points on the boundary of the polygon $\mathcal{P}$. An example of a set of fix-points computed for a sample polygon is presented in Fig.~\ref{fig:polyline_simplification_overview} (black circles).
The polylines considered for simplification include all borders between the fix-points that are not part of the boundary of $\mathcal{P}$ (black-red circles in the figure). The \emph{simplifyBorders} procedure applies the k-point polyline simplification algorithm to these borders.

The simplification algorithm applied to each border should fulfill two key requirements. First, the resulting polyline should have minimal influence on the compactness score of each sub-polygon (\emph{compactness requirement}). Second, the effects of the border simplification on the area size of each sub-polygon should be minimized (\emph{area requirement}).

In~\cite{BOSE2006554}, the authors proved that the problem of approximating a polyline by a k-point polyline that fulfills the compactness and area requirements is an NP-hard problem.
Additionally, a solution limited to the case of an x-monotone path respecting the area requirement was proposed.
Most polyline simplification algorithms, such as variants of the Ramer$-$Douglas$-$Peucker algorithm~\cite{RAMER1972244}\cite{doi:10.3138/FM57-6770-U75U-7727}, address the compactness requirement by ensuring that the distance between the simplified polyline is within a threshold of the original polyline. Thus, the approximated polyline has similar shape to the original polyline and as a consequence the compactness of the polygons defined by either the approximated or original polylines remains similar. 
In this work, the proposed k-point simplification algorithm adopts this idea. The exact relation between the distance measure threshold values and the compactness score is left to be investigated in the future work.

The proposed method incrementally tries to find a k-point polyline that fulfills both area and compactness requirements with increasing number of $k$ points, starting with $k=3$.
For each value of $k$, a gradient descent~\cite{gradDesc-ruder2016overview} method is used to find positions of the intermediate points in the new k-polyline, where the objective is to satisfy the area requirement.
The objective function used in the optimization is based on Green's theorem~\cite{citeulike:3033145} which allows for a calculation of areas enclosed by the original border and its k-point simplifying polyline. 
After the optimization, the algorithm checks if the maximum distance between the new k-polyline and the original border is within a given threshold, that is, if the compactness requirement is satisfied. 
If the distance is larger than the threshold, the value of $k$ is incremented, and the procedure is repeated.


\begin{figure}[!t]
\centering
\includegraphics[width=0.7\columnwidth]{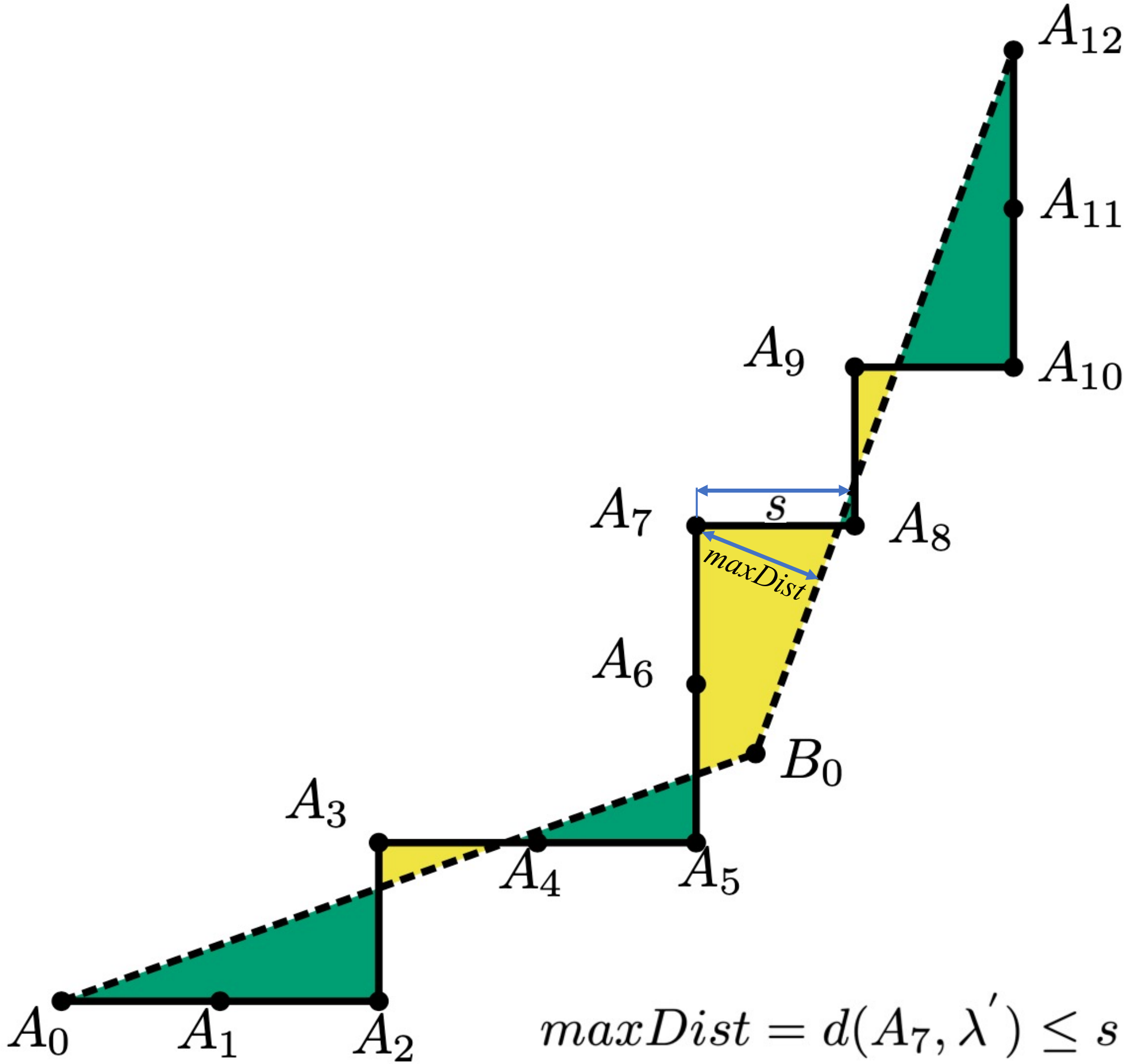}
\caption{
An example of a border simplification by a k-point polyline. The border polyline defined by $\PL = \{A_0$,\ldots,$A_{12}\}$ is simplified by a new 3-point polyline $\PLS = \{A_0,B_0,A_{12}\}$ satisfying two critical requirements: 1) maintain the same area between the two sides of the polyline $\PLS$ (i.e. the sum of yellow and the sum of  green areas are equal), 2) the new polyline $\PLS$ has similar shape to the original polyline $\PL$ which in turn minimizes the influence of the simplification on the compactness of sub-polygons located on both sides of the border.} 
\label{fig:polyline_simplification}
\end{figure}

Before describing the \emph{simplifyBorders} algorithm in detail, the following notation is introduced. 
Let $\PL$ denote the input polyline defined by a set of $m$ points $\{A_0, \ldots, A_m\}$. A k-point polyline that simplifies $\PL$ is denoted as $\PLS$.
Note, the first and the last points in  $\PLS$ are $A_0$ and $A_m$, respectively. Thus, the goal of the algorithm is to find positions of $l=k-2$ intermediate points $\{B_0...B_{l-1}\}$ to complete the polyline.
An example of single polyline simplification is shown in  Fig.~\ref{fig:polyline_simplification}. The original polyline $\PL$ consists of 13 points $\{A_0,\ldots,A_{12}\}$ and its simplified by a 3-point polyline $\PLS$ defined by three points $\{A_0, B_0,A_{12}\}$. The area requirement is satisfied as the areas between the two sides of the polyline $\PLS$ enclosed by $\PL$ are equal (i.e. the sum of the yellow and the sum of green areas are equal). The maximum distance between the points of border $\PL$ and the simplifying polyline $\PLS$ is below a given threshold satisfying the compactness requirement.


\begin{algorithm}
	\DontPrintSemicolon
	\KwIn{$\MP,\left<\mathcal{P}_1, \ldots, \mathcal{P}_n\right>, s$}
	\KwOut{$\left<\mathcal{P}_1, \ldots, \mathcal{P}_n\right>$}
	\Parameter{\textit{MaxIter$_{SB}$}}
	$\PLSET \gets$ \emph{getBordersToSimplify}$(\left<\mathcal{P}_1, \ldots, \mathcal{P}_n\right>)$\;
	\For{$\PL$ \textup{\textbf{in}} $\PLSET$} {
	    \emph{Finished} $\gets false$\;
	    $l \gets 1$\;
	    \While{$\neg$Finished} {
	        $ \{B_0,\ldots,B_{l-1}\} \gets$ \emph{initPoints}$(\PL, l)$\;
	        $\PLS \gets \{A_0,B_0,\ldots,B_{l-1},A_m\}$\;
	        $\PLS \gets$ \emph{gradDesc}$(\PL,\PLS,$ \emph{MaxIter}$_{SB})$\;
            \emph{maxDist} $\gets$ \emph{calcMaxDist}$(\PL,\PLS)$\;
	        \lIf { maxDist $ > s $}
		    {
		        $l \gets l+1$
		    } \Else{$\PL \gets \PLS$\;
		        \emph{Finished} $\gets true$}
	    }    
	}
	$\langle \MP_1,\ldots,\MP_n\rangle \gets$ \emph{updateBorders}$(\PLSET)$\;
	\Return{$\langle \MP_1,\ldots,\MP_n\rangle$}\;
	\caption{{\sc SimplifyBorders}}
	\label{algo:SimplifyBoundaries}
\end{algorithm}

The \emph{SimplifyBorders} algorithm is presented in~\Cref{algo:SimplifyBoundaries}.
The input to the algorithm includes the original polygon $\MP$, the set of sub-polygons $\langle\MP_1,\ldots,\MP_n \rangle$ and the size of the grid-cell $s$ (\cref{eq:grid_cell_size}) used as a threshold for satisfying the compactness requirement. The parameter $MaxIter_{SB}$ defines the maximum number of iterations of the gradient descent algorithm used to find the intermediate point positions of the simplifying k-point polyline $\PLS$.

The \emph{SimplifyBorders} algorithm starts with identifying a set of borders $\PLSET$ that need to be simplified (Line 1). For each border $\PL \in \PLSET$ a k-point simplifying polyline $\PLS$ is calculated in Lines 2-13. The number of intermediate points $l$ in $\PLS$ is initialized to 1 (Line 4). The main optimization loop for finding the positions of intermediate points of $\PLS$ is executed in Lines 5-13. 

The optimization loop starts with the initialization of intermediate points $\{B_0,\ldots,B_{l-1}\}$ (Line 6). The $initPoints(\PL, l)$ function returns a list of $l$ points equally spaced along the original polyline $\PL$. The initial k-point polyline $\PLS$ is constructed by combining the start ($A_0$) and end point ($A_m$) of $\PL$ with the set of intermediate points $\{B_0,\ldots,B_{l-1}\}$ (Line 7).

The gradient descent optimization method ($gradDesc$, Line 8) is used to calculate the positions of intermediate points of $\PLS$.
For a given set of $l$ points $\{B_0,\ldots,B_{l-1}\}$ the method solves the following optimization problem:

\begin{align}
  \min\limits_{B_0,\ldots,B_{l-1}} \mid green(A_0,\ldots,A_i,\ldots,A_m, B_{l-1},\ldots, B_0)\mid
\end{align}

\noindent where $green$ function is based on Green's theorem \cite{citeulike:3033145} and allows for calculation of areas enclosed by $\PL$ and $\PLS$. 
If the sum of areas on each side of $A_0,\ldots,A_i,\ldots,A_m$ enclosed by $B_{l-1},\ldots,B_0$ are equal then $green(A_0,\ldots,A_i,\ldots,A_m, B_{l-1},\ldots,B_0) = 0$. The green formula is defined as:

\begin{equation}
  green((x_0, y_0),\ldots,(x_n,y_n)) = \sum\limits_{i=0}^n (x_iy_{i+1} - x_{i+1}y_i)
\end{equation}

\noindent where $(x_{n+1},y_{n+1}) = (x_0, y_0)$ and $n=m+l$.


Minimizing the value of the green formula when choosing the positions of the intermediate points of $\PLS$ ensures that the area requirement is satisfied.

The resulting configuration of the k-point polyline $\PLS$ is evaluated using the distance measure to check if it satisfies the compactness requirements (Line 9-13). The $calcMaxDist$ function computes the maximum distance between the points of $A_1,\ldots,A_{m-1}$ and $A_0, B_0,\ldots,B_{l-1}, A_m$ according to:
\begin{equation}
    maxDist = \max\limits_{i=1...m-1} d(A_i, \PLS) 
\end{equation}
An example result of $maxDist$ calculation for a sample polyline is depicted  in Fig.~\ref{fig:polyline_simplification}.
If the value of $maxDist$ is larger than the threshold defined as the grid-cell size $s$ (\cref{eq:grid_cell_size}), the compactness requirement is not satisfied and the optimization loop (Lines 5-13) is repeated with an increased number of intermediate points $l \gets l + 1$. Otherwise, the newly calculated k-point polyline $\PLS$ meets both area and compactness requirements, and the main optimization loop is terminated (Line 13) and continues its execution for the remaining borders in $\PLSET$.

Finally, once the process of finding simplifying k-point polylines for all borders in $\PLSET$ is finished, the algorithm assigns newly calculated polylines to respected sub-polygons in $\langle \MP_1,\ldots,\MP_n\rangle$ (Line 14).


\section{Empirical Evaluation}\label{sec:evaluation}
The proposed \emph{AreaDecompose} algorithm has been evaluated in a series of runs with different configurations applied to a set of randomly generated polygons. The generated set consist of 200 non-convex polygons\footnote{\url{https://gitlab.liu.se/lrs/pad_polygons}}.

Several compactness scores exist in the literature that can be used in the optimization phase of the \emph{AreaDecompose} algorithm. For the experimental evaluation, a set of the five most commonly used compactness metrics has been chosen. 
The first metric is the \emph{Schwartzberg score}~\cite{schwartzberg65}. The score is computed as the ratio between the perimeter of the input polygon $\MP$ and the circumference of a circle which has the same area size:
\begin{equation}
\centering
    S=\frac{2 \cdot \sqrt{\pi \cdot \MA(\MP)}}{P_\MP}  
\end{equation}
\noindent where $P_\MP$ is the perimeter of the polygon $\MP$.

The second metric is the \emph{Polsby–--Popper score}~\cite{polsby:91}, which is calculated as a ratio of the $\MP$ polygon area and a circle area whose circumference is equal to the perimeter of the polygon:
\begin{equation}
\centering
PP=\frac{4\cdot \pi \cdot \mathcal{A}(\mathcal{P})}{P^2_\mathcal{P}}  
\end{equation}

The third metric is the \emph{Reock degree of compactness}~\cite{Reock:61}. It is computed as the ratio between the polygon area and the area of the smallest circle which encloses it:
\begin{equation}
\centering
R=\frac{\mathcal{A}(\mathcal{P})}{A_{MBC}}  
\end{equation}
The fourth compactness metric is the \emph{two balls score} derived from the definition of the $\alpha$-fat property~\cite{EFRAT2000215}. The score is computed as the ratio between the circumference of the maximum inscribed circle $s^+$ in the polygon $\MP$ and the circumference of the minimum enclosing circle $s^-$ of the polygon $\MP$:
\begin{equation}
\centering
TB=\frac{\mathcal{C}(s^+)}{\mathcal{C}(s^-)}
\end{equation}
\noindent where $\mathcal{C}(s^+)$ and $\mathcal{C}(s^-)$ are circumferences of the inscribed and enclosing circles, respectively.



The final metric used in the evaluation is the length-to-width ratio of the rotated minimum bounding rectangle (RMBR) of the polygon $\MP$:
\begin{equation}
\centering
LW=\frac{W_{RMBR}}{L_{RMBR}}
\end{equation}
\noindent where $W_{RMBR}$ and $L_{RMBR}$ is the width and length of the RMBR.

The values for all compactness scores range from 0 to 1, where values closer to 1 indicate more compact polygons. The Schwartzberg score was used in the objective function (\cref{eq:objective2}) whenever the optimization methods in the \emph{AreaDecompose} algorithm were utilized. Other compactness scores listed above were used for the evaluation of results. A \emph{Collective} compactness score was calculated as an average of all scores for each generated result.

The evaluation of the algorithms was performed on a set of randomly generated polygons. The problem parameters used for different runs are summarized in Table~\ref{tab:experimental_setup}. The algorithms were applied to five different problems (categorized as Case 1-5) with varying the number of sub-polygons and their sizes. In the simplest case (Case 1), the polygons were divided into two equally sized sub-polygons. The most complex case (Case 5) involved dividing polygons into a randomly selected number of sub-polygons (up to 9) with their sizes selected randomly. Cases 1-4 were applied once on the polygon dataset, while Case 5 was applied twice.

\begin{table}[]
\centering
\begin{tabular}{|c|c|c|c|c|}
\hline
 Case\# & \#Polygons & \#Partitions & Weights $\Omega$   \\ \hline
  1 & 200 & 2 & 0.5, 0.5   \\ \hline
  2 & 200 & 3 & 0.166, 0.333, 0.5 \\ \hline
  3 & 200 & 4 & 0.1, 0.2, 0.4, 0.5 \\ \hline
  4 & 200 & 5 & 0.2, 0.2, 0.2, 0.2, 0.2   \\ \hline
  5 & 400 & RND (2,9) & RND   \\ \hline
 \end{tabular}
 \caption{Problem configurations (Cases) used in the experimental evaluation.}
 \label{tab:experimental_setup}
\end{table}

Five configurations of the \emph{AreaDecompose} algorithm were used. The first three configurations included running a single optimization algorithm. These included running either the heuristic-based PFH algorithm (Section~\ref{sec:pfh}), the CMA-ES (Section~\ref{sec:CMAES}) or the Random Search (RS, Section~\ref{sec:RS}). The remaining two configurations involved using a combination of the heuristic-based PFH with one of the other remaining optimization methods (i.e. CMA-ES and RS). The results generated by the heuristic-based algorithm were used as initial solutions for the optimization methods. Additionally,  three different tolerance values for area size errors were used (i.e. $\tau=\{1\%, 5\%, 10\%\}$).

Sub-polygons generated by the \emph{AreaDecompose} algorithm when solving Cases 1-5 were compared to the polygons generated by the \emph{Hert\&Lumelsky} algorithm~\cite{Hert1998PolygonAD} using the same input parameters.
The experiments were run using a computer equipped with an AMD Ryzen 9 3900XT 12-Core Processor utilizing a single core. The \emph{Hert\&Lumelsky} algorithm was implemented in Python, while the \emph{AreaDecompose} was implemented in C++ programming language.

\begin{figure}[!t]
    \centering
    \includegraphics[width=1.0\columnwidth]{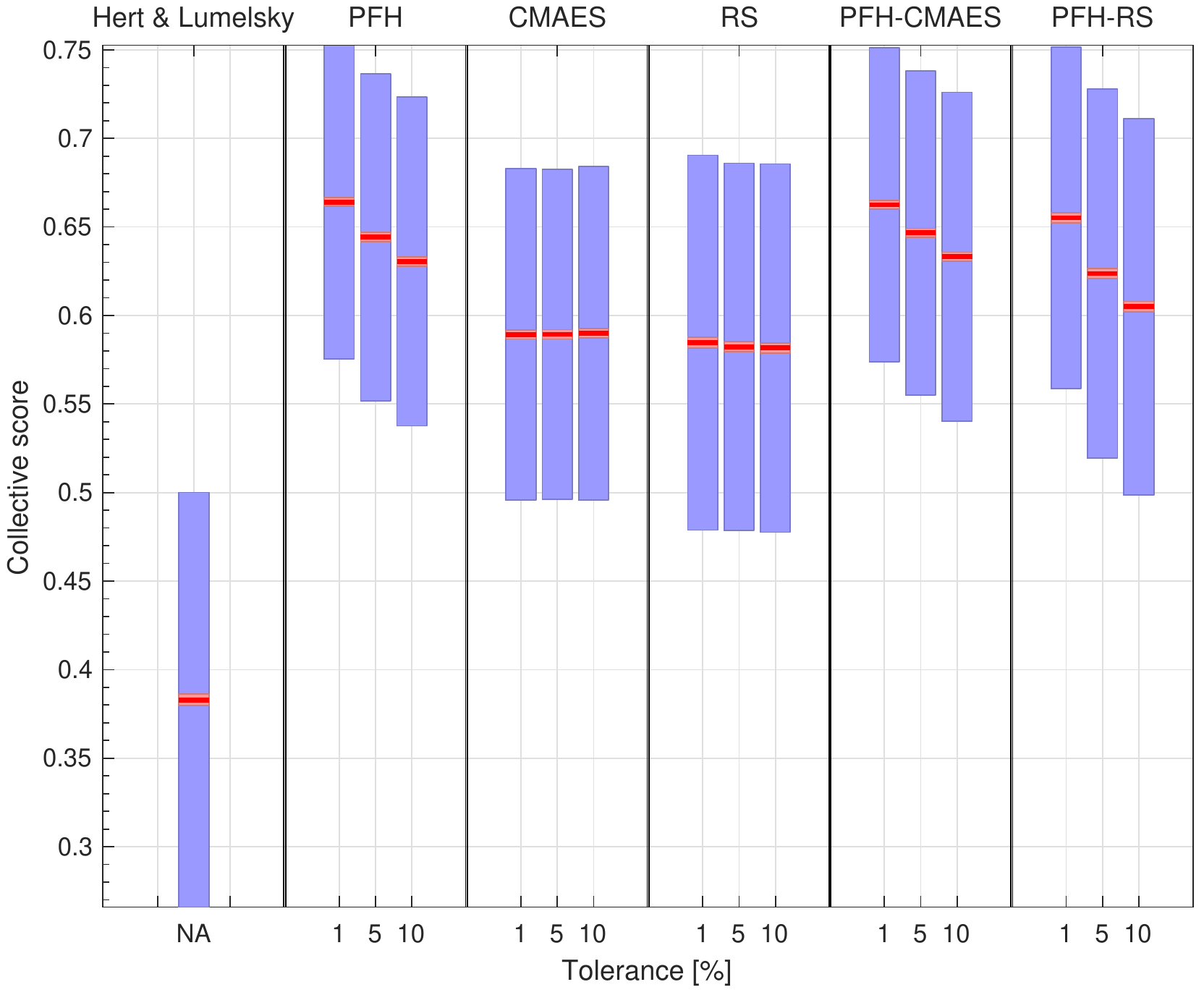}
    \caption{Collective compactness scores for different algorithm configurations calculated as an average values over all cases.}
    \label{fig:results_combined_score}
\end{figure}

\begin{figure}[!t]
    \centering
    \includegraphics[width=1.0\columnwidth]{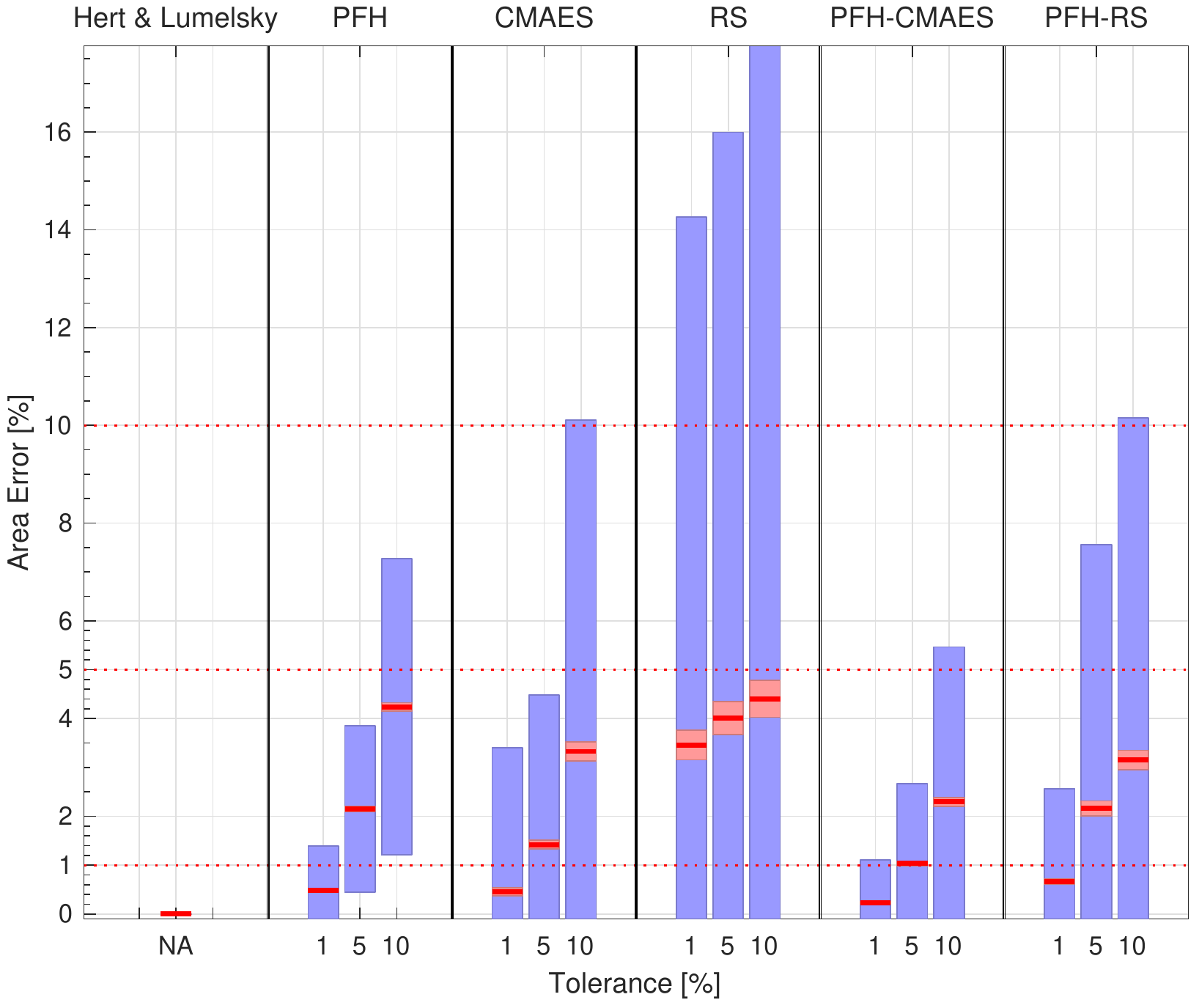}
    \caption{Area size errors for different algorithm configurations calculated as an average values over all cases.}
    \label{fig:results_combined_error}
\end{figure}

\begin{figure}[!t]
    \centering
    \includegraphics[width=1.0\columnwidth]{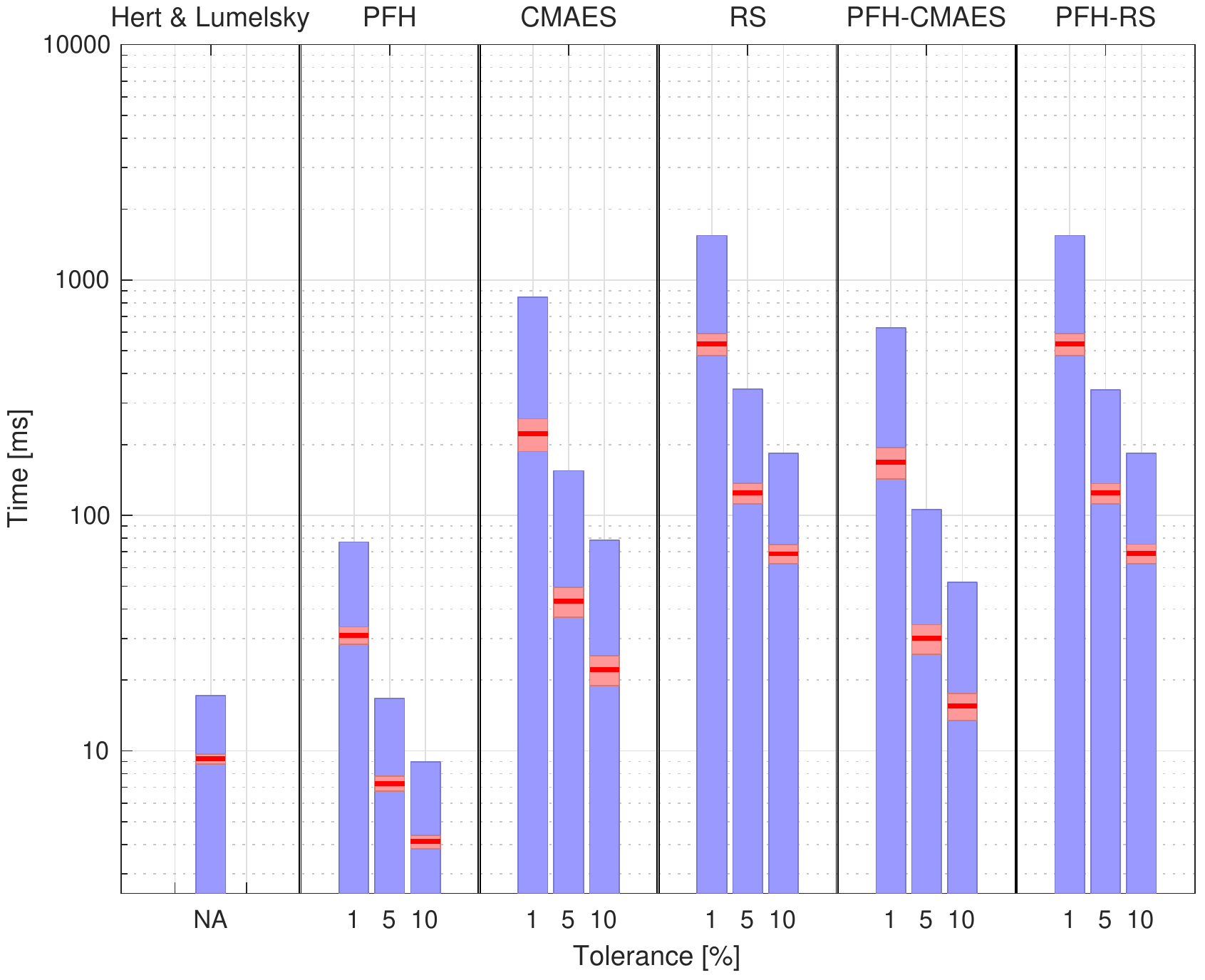}
    \caption{Running times for different algorithm configurations shown in log-scale calculated as an average values over all cases.}
    \label{fig:results_combined_time}
\end{figure}

The results of the experimental evaluations for different algorithm configurations and tolerance values $\tau$ were computed as an average value over all cases considering three metrics. The metrics included the quality and accuracy of the solutions and the efficiency of the algorithm.
The quality of the solutions measured as \emph{Collective} compactness scores for the generated sub-polygons is presented in Fig.~\ref{fig:results_combined_score}.
The sub-polygon area errors representing the accuracy of the solutions are depicted in Fig.~\ref{fig:results_combined_error}.
Finally, the algorithm efficiency measured as the algorithm execution times are depicted in Fig.~\ref{fig:results_combined_time}.

The \emph{Hert\&Lumelsky} algorithm generated sub-polygons with the lowest compactness score (0.38) among all the algorithms considered (Fig.~\ref{fig:results_combined_score}). In comparison, the \emph{AreaDecompose} algorithm with its five different configurations provided solutions with sub-polygons' compactness scores between 0.58 to 0.66. The worst performing configurations included running only the CMA-ES or RS optimization methods. These optimization methods do not perform well when searching for an optimal solution given the initial parameters of the potential field configurations. The optimization methods most likely find sub-optimal solutions at local minima. Additionally, the three different tolerance values for area size errors ($\tau$) did not influence the achieved compactness scores.
The heuristics-based PFH and its combinations with CMA-ES and RS generated results with the highest compactness scores ranging between 0.61 and 0.66. In these configurations, lower tolerance value settings ($\tau$) led to solutions with higher compactness scores. This is expected as the lower $\tau$ values will result in a higher number of grid cells used, thus allowing to find solutions with more compact sub-polygons.

The average area size error for solutions generated by the \emph{Hert\&Lumelsky} algorithm was recorded at 3.7e-4\% (Fig.~\ref{fig:results_combined_error}). This is as expected since the algorithm guarantees the area sizes of sub-polygons to be exact. In case of the \emph{AreaDecomse} algorithm, the area size errors are bounded by the tolerance parameter $\tau$. All configurations of the algorithm generated solutions with an error below the value of the tolerance $\tau$, except the RS. The most accurate solutions were generated by the PFH-CMAES configuration with errors as low as 0.23\% for the case of $\tau=1\%$.

The \emph{AreaDecompose} algorithm efficiency measured as its execution times varied depending on the chosen configuration and the tolerance of the area size error parameter. This is as expected since the execution times depend on the number of grid cells used while searching for a solution. Overall, the algorithm execution times varied between 4.1ms (PFH, $\tau=10\%$) and 535ms (RS, $\tau=1\%$). Execution times for the \emph{Hert\&Lumelsky} algorithm were measured at 9.2ms, which was higher than in the case of the PFH run with $\tau=5\%$ (4.1ms) and $\tau=10\%$ (7.3ms). Other \emph{AreaDecompose} configurations resulted in higher execution times especially for configurations were the heuristic-based PFH was combined with the CMA-ES or the RS techniques.

Overall, the PFH and PFH combined with CMA-ES were the two best \emph{AreaDecompose} algorithm configurations when considering all three measured aspects. Results comparing these two configurations to the \emph{Hert\&Lumelsky} algorithm are presented in Table~\ref{tab:experimental_results_summary}. For the case of the area size error $\tau=1\%$ the PFH and its combination with CMA-ES generated solutions with a $73\%$ higher collective compactness score in comparison to the \emph{Hert\&Lumelsky} algorithm. The PFH-CMAES configuration provided solutions with significantly lower area size error (0.23\%) when compared to the PFH configuration (0.48\%) at a cost of a higher computational time required (31ms vs 168ms).

\begin{figure}[!t]
    \centering
    \includegraphics[width=1.0\columnwidth]{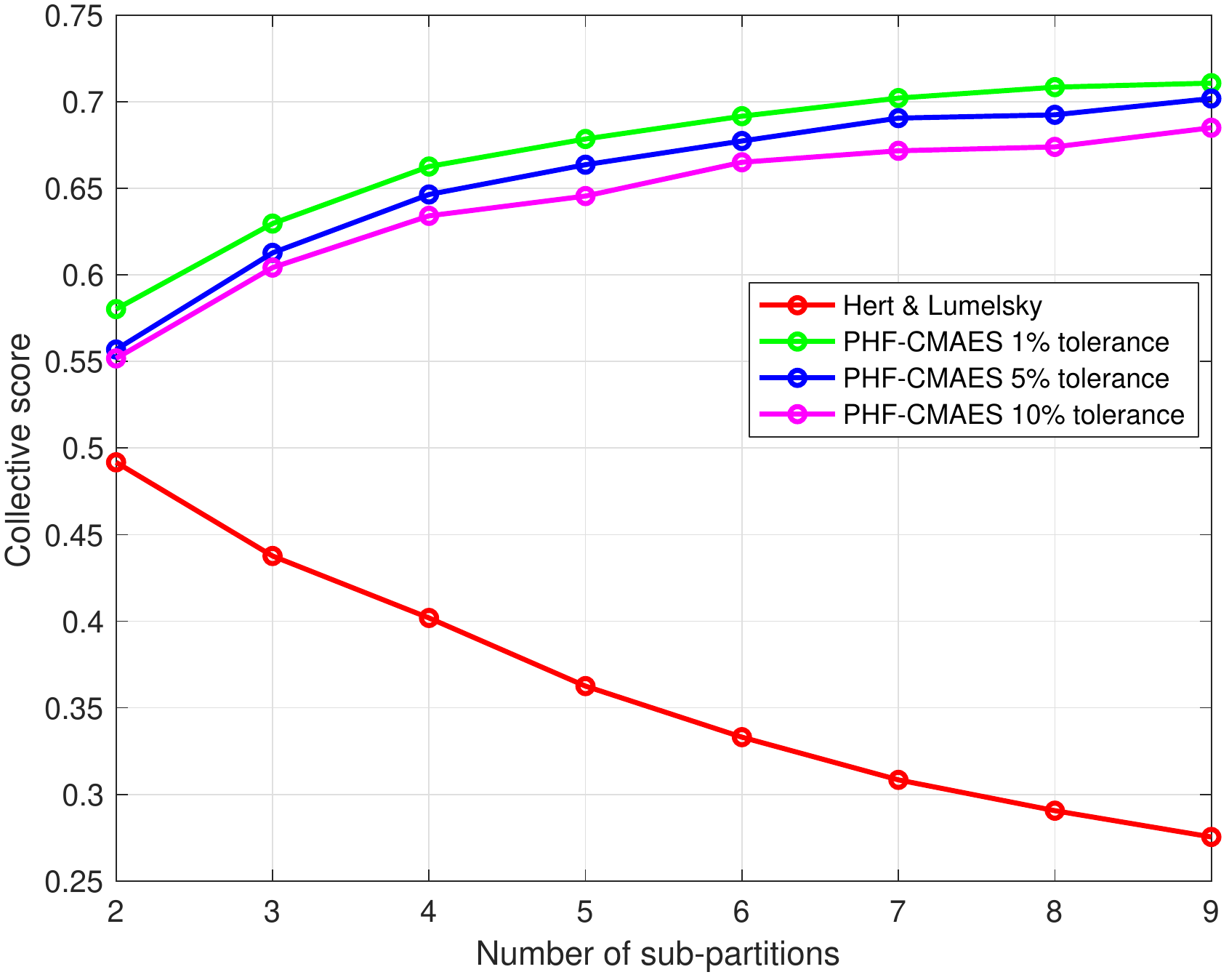}
    \caption{Dependency between the average collective score and the number of sub-polygons for different algorithm configurations.}
    \label{fig:results_combined_score_vs_subpolygons}
\end{figure}


An interesting dependency between the number of sub-polygons requested in the problem and the compactness scores has been observed. This is especially visible in problems where the original polygon is non-convex, such as the polygons included in the evaluation dataset. Dividing a non-convex polygon into a small number of partitions (e.g. two) is restricted by the original polygon borders. More compact sub-polygons can be generated when the problem involves dividing the polygon into a larger number of sub-polygons. 
Fig.~\ref{fig:results_combined_score_vs_subpolygons} presents the average collective compactness scores as a function of the number of sub-polygons for different algorithm configurations. 
Interestingly, the \emph{Hert\&Lumelsky} algorithm generally generates less compact sub-polygons as the number of sub-partitions grows. This is attributed to the divide-and-conquer technique used in the algorithm, and the results obtained in our evaluation confirm the findings discussed in~\cite{Hert1998PolygonAD}. In the case of the \emph{AreaDecompose} algorithm, this trend is reversed. That is, the higher the number of sub-polygons requested in the problem, the higher their compactness.
The PFH-CMAES configuration of the \emph{AreaDecompose} algorithm generated sub-polygons with compactness scores between 0.55 (2 sub-polygons) up to 0.71 (9 sub-polygons). The \emph{Hert\&Lumelsky} algorithm found solutions with compactness scores between  0.27 (9 partitions) and 0.49 (2 sub-polygons). While the PFH-CMAES outperforms the \emph{Hert\&Lumelsky} algorithm when considering the sub-polygon compactness, the difference becomes much more significant as the number of sub-polygons requested in the problem increases. In the extreme case of 9 sub-polygons, the PFH-CMAES generated sub-polygons which are 258\% more compact than the results provided by the \emph{Hert\&Lumelsky} algorithm.

\begin{table}[]
\centering
\begin{tabular}{|c|c|c|c|c|}
\hline
Algorithm      & Tolerance  & Collective  &  Area Error  & Time \\ 
               & $\tau$     & Score       &   [\%]       &  [ms] \\ \hhline{|=|=|=|=|=|}
	           & 1\%        & 0.66        & 0.23         & 168.45 \\
PFH-CMAES      & 5\%        & 0.65        & 1.03         & 30.02 \\
               & 10\%       & 0.63        & 2.29         & 15.48 \\ \hline
               & 1\%        & 0.66        & 0.48         & 31.02 \\
PFH	           & 5\%        & 0.64        & 2.14         & 7.26 \\
	           & 10\%       & 0.63        & 4.24         & 4.11 \\ \hline
Hert\&Lumelsky & NA         & 0.38        & $\approx$ 0       & 9.24 \\ \hline
 \end{tabular}
 \caption{Comparison of the results for the two best AreaDecompose algorithm configurations and Hert\&Lumelsky approach.}
 \label{tab:experimental_results_summary}
\end{table}



\subsection{Other Application Examples}\label{seq:redistricting}

\begin{figure*}
    \centering
    \includegraphics[width=0.48\textwidth]{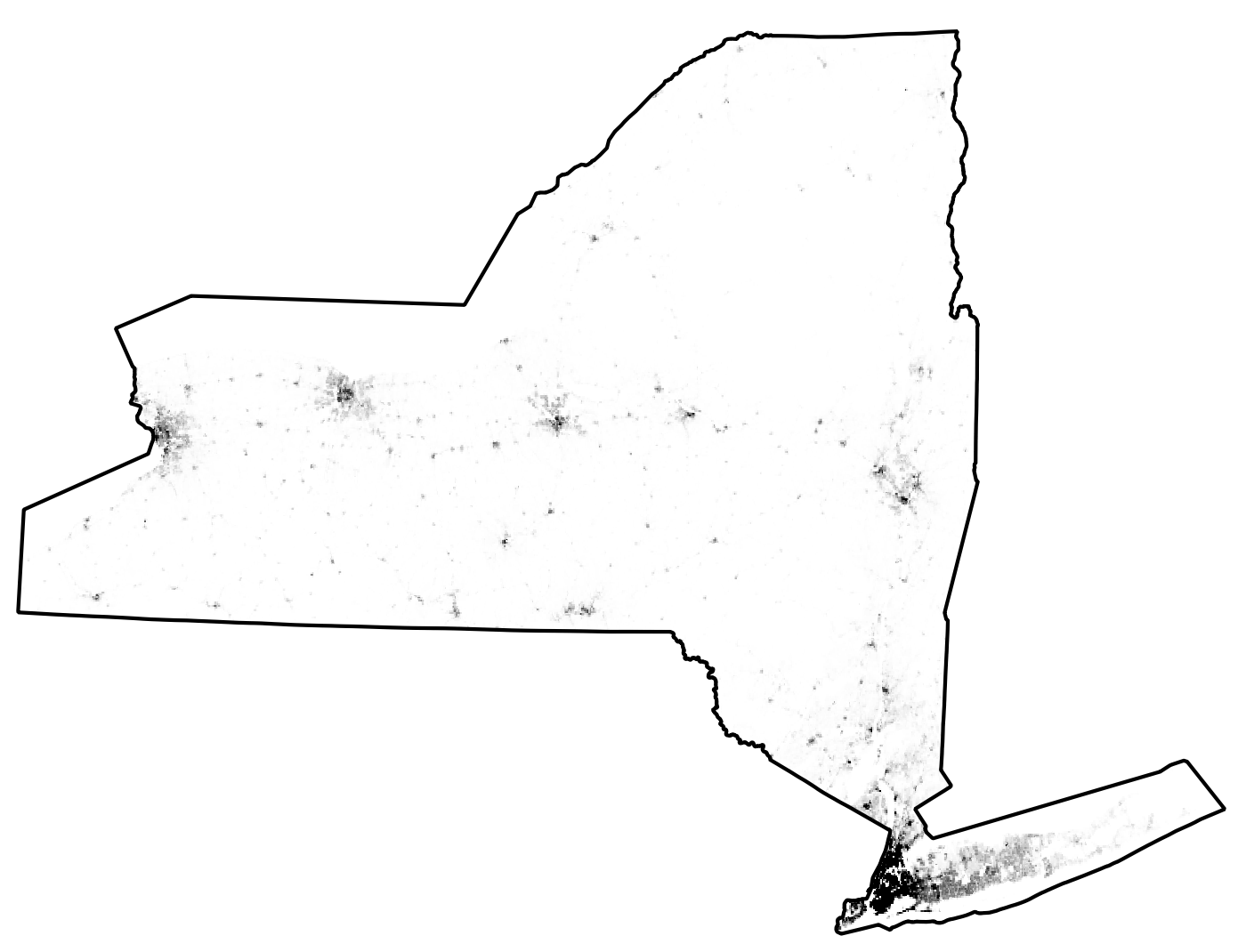}
    \includegraphics[width=0.48\textwidth]{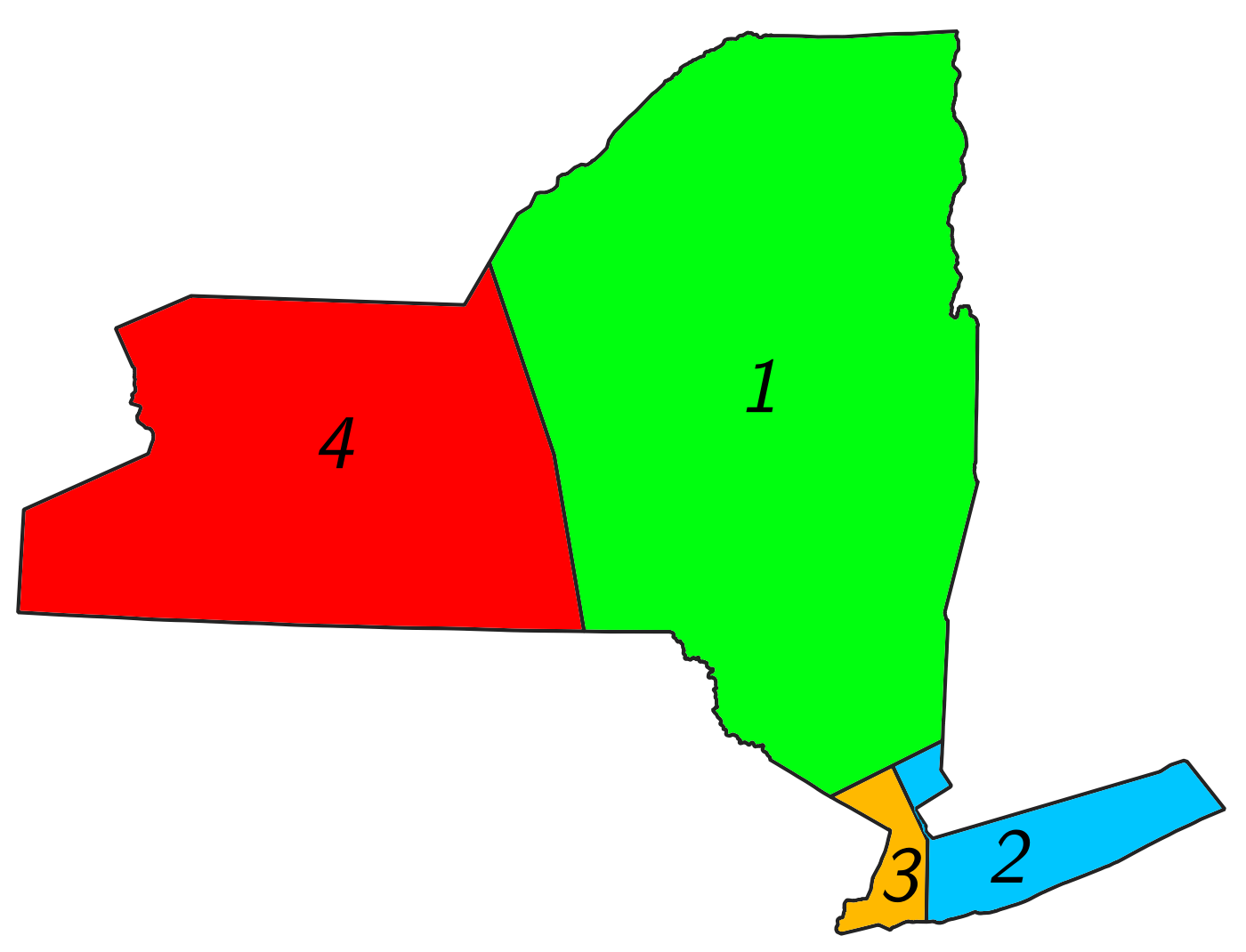}
    \caption{The population density in the state of New York is depicted on the left, where the white color represents the lowest density and black the highest.
    Result of the division of the state of New York based on population density is depicted on the right.}
    \label{fig:ny}
\end{figure*}

The proposed problem representation that uses a grid cell discretization and potential field model can be utilized for solving other classes of problems than area partitioning. One such class includes GIS related problems where the spacial data is represented as \emph{rasters}.
A raster consists of a two dimensional matrix of cells (or pixels) where each cell contains a value representing information. Rasters are commonly used in GIS systems to represent spacial data due to their simple format and ability to visualize them in the form of an image or a map. An example is a population density map, where each pixel of the image contains the population size~\cite{langford1994generating}.
The grid-based representation used as the basis of \emph{AreaDecompose} allows for a straight forward association between the value contained in a raster and a cell of the grid used in the process of partitioning.
By calculating the population of each grid cell $\nu$ based on the population density map, the \emph{AreaDecompose} algorithm can be used to solve the problem of dividing a polygon based on the population density instead of the total surface of each partition.
In that case, $\mathcal{A}(\mathcal{P}_i)$ introduced in \cref{sec:algorithm} would not refer to the total surface, but instead, to the total population living in $\mathcal{P}_i$.

To demonstrate this application, the \emph{AreaDecompose} is applied to the following redistricting problem. Given the population density map of the state of New York, calculate four districts with an equal population. The input population density map and districts generated by the \emph{AreaDecompose} algorithm are shown in Fig.~\ref{fig:ny}. \Cref{table:ny} presents statistics for the resulting sub-regions, including the population error and three compactness scores. 
Each generated sub-region, with the exception of sub-region 2, maintains high compactness scores. In the case of region 2, the partitioning is constrained by the border of the state.
Based on the input population density data, each district should ideally include 5.053 million inhabitants. Districts generated by the \emph{AreaDecompose} algorithm have a population error of at most $2.77\%$ while maintaining high compactness scores.


\begin{table}[]
\centering
\begin{tabular}{|c|c|c|c|c|c|c|}
\hline
 Sub-    & Popu-  & Pop. & Surface& Schwarz- & Reock & Polsby- \\ 
  region & lation & Error& [$km^2$] & berg &  & Popper \\ 
         & [M]    & [\%] &          &      &  &   \\ \hline
  1 & 5.114 & 1.19 & 84,052 & 0.79 & 0.56 & 0.62  \\ \hline
  2 & 5.078 & 0.49 & 7,715 & 0.56 & 0.27 & 0.31 \\ \hline
  3 & 5.104 & 1.00 & 2,694 & 0.62 & 0.35 & 0.39 \\ \hline
  4 & 4.917 & 2.77 & 46,849 & 0.79 & 0.51 & 0.62 \\ \hline
 \end{tabular}
 \caption{Results of partitioning of the New-York state based on spacial population density data.}
 \label{table:ny}
\end{table}


\section{Conclusion and future work}\label{sec:conclusions}

In this paper, we have considered the problem of area partitioning, which deals with the division of non-convex polygons into a set of non-overlapping sub-polygons with defined area sizes.

The area partitioning problem arises in terrain covering applications in collaborative robotics, where the goal is to find a set of efficient plans for a team of heterogeneous robots to cover a given area. 
In this application context, the shape characteristics of the sub-regions generated as a result of solving the area partitioning problem can improve the efficiency with which robotic tasks can be performed within each sub-region. In fact, maximizing the compactness of sub-polygons directly influences the optimality of any generated motion plan. The state-of-the-art approach used in this application and proposed by Hert\&Lumelsky in~\cite{Hert1998PolygonAD} provides an algorithm that guarantees the satisfaction of area size constraints, but does not consider sub-region shape constraints. The algorithms that have been presented in this paper focus on addressing both the size and shape constraints while solving the area partitioning problem.

In this work, an extended formulation of the problem has been proposed that in addition to respecting area size constraints, aims at maximizing compactness scores of the generated sub-polygons.
The proposed problem formulation is based on grid cell discretization and a potential field model and allows for applying general optimization algorithms.
An algorithm, the \emph{AreaDecompose}, has been presented for solving the extended area partitioning problem efficiently.
The algorithm combines a new heuristic-based method with existing optimization techniques and two post-processing methods.


An empirical evaluation has been presented comparing the error on the area size and the compactness metric when using different configurations of the proposed partitioning algorithm.  In addition, a comparative analysis is made between the algorithms proposed in this paper and the state-of-the-art approach proposed by Hert\&Lumelsky in~\cite{Hert1998PolygonAD}.
The results have shown that the new methods proposed in this paper can generate on average up to 73\% more compact partitions than the one from Hert\&Lumelsky, while remaining within a specified tolerance on the error of the size of the partitions.
Additionally, the best results were obtained when the \emph{AreaDecompose} algorithm was configured to use a combination of heuristics-based PFH and the CMA-ES optimization technique. The PFH quickly finds initial solutions, which are then further improved using the CMA-ES optimization, reducing the errors on the area constraints and improving the compactness of each sub-polygon.

While the standard area partitioning problem has been mainly defined to be used in terrain covering applications in robotics, the extended problem definition proposed here can be easily adopted to other GIS-related applications. An example of such an application, also presented in the paper, is the problem of redistricting for the state of New York.


In this work, the polygon simplification step aims to ensure that the areas of the generated sub-polygons do not change and to keep the simplified polygon within bounds of the original polygon to maintain its compactness properties.
In future work, it would be interesting to investigate if the simplification step can be used to reduce the error on the area constraint and also improve the compactness property.

\bibliographystyle{IEEEtran}
\bibliography{biblio_new}

\begin{thebibliography}{10}
\providecommand{\url}[1]{#1}
\csname url@samestyle\endcsname
\providecommand{\newblock}{\relax}
\providecommand{\bibinfo}[2]{#2}
\providecommand{\BIBentrySTDinterwordspacing}{\spaceskip=0pt\relax}
\providecommand{\BIBentryALTinterwordstretchfactor}{4}
\providecommand{\BIBentryALTinterwordspacing}{\spaceskip=\fontdimen2\font plus
\BIBentryALTinterwordstretchfactor\fontdimen3\font minus
  \fontdimen4\font\relax}
\providecommand{\BIBforeignlanguage}[2]{{%
\expandafter\ifx\csname l@#1\endcsname\relax
\typeout{** WARNING: IEEEtran.bst: No hyphenation pattern has been}%
\typeout{** loaded for the language `#1'. Using the pattern for}%
\typeout{** the default language instead.}%
\else
\language=\csname l@#1\endcsname
\fi
#2}}
\providecommand{\BIBdecl}{\relax}
\BIBdecl

\bibitem{Asano:83}
T.~{Asano} and T.~{Asano}, ``Minimum partition of polygonal regions into
  trapezoids,'' in \emph{24th Annual Symposium on Foundations of Computer
  Science (sfcs 1983)}, Nov 1983, pp. 233--241.

\bibitem{Asano:86}
\BIBentryALTinterwordspacing
T.~Asano, T.~Asano, and H.~Imai, ``Partitioning a polygonal region into
  trapezoids,'' \emph{J. ACM}, vol.~33, no.~2, pp. 290--312, Apr. 1986.
  [Online]. Available: \url{http://doi.acm.org/10.1145/5383.5387}
\BIBentrySTDinterwordspacing

\bibitem{feng1975decomposition}
H.-Y. Feng and T.~Pavlidis, ``Decomposition of polygons into simpler
  components: Feature generation for syntactic pattern recognition,''
  \emph{IEEE Transactions on Computers}, vol. 100, no.~6, pp. 636--650, 1975.

\bibitem{moitra1991finding}
D.~Moitra, ``Finding a minimal cover for binary images: An optimal parallel
  algorithm,'' \emph{Algorithmica}, vol.~6, no.~1, pp. 624--657, 1991.

\bibitem{lodi1978two}
E.~Lodi, F.~Luccio, C.~Mugnai, and L.~Pagli, ``On two-dimensional data
  organization i,'' \emph{Fundamenta Informaticae}, vol.~2, no.~1, pp.
  211--226, 1978.

\bibitem{area-constr-christou1996optimal}
I.~T. Christou and R.~R. Meyer, ``Optimal equi-partition of rectangular domains
  for parallel computation,'' \emph{Journal of Global Optimization}, vol.~8,
  no.~1, pp. 15--34, 1996.

\bibitem{Hert1998PolygonAD}
S.~Hert and V.~J. Lumelsky, ``Polygon area decomposition for multiple-robot
  workspace division,'' \emph{Int. J. Comput. Geometry Appl.}, vol.~8, pp.
  437--466, 1998.

\bibitem{Skorobogatov2021}
\BIBentryALTinterwordspacing
G.~Skorobogatov, C.~Barrado, E.~Salam\'{i}, and E.~Pastor, ``Flight planning in
  multi-unmanned aerial vehicle systems: Nonconvex polygon area decomposition
  and trajectory assignment,'' \emph{International Journal of Advanced Robotic
  Systems}, vol.~18, no.~1, p. 1729881421989551, 2021. [Online]. Available:
  \url{https://doi.org/10.1177/1729881421989551}
\BIBentrySTDinterwordspacing

\bibitem{li2013}
W.~Li, M.~F. Goodchild, and R.~Church, ``An efficient measure of compactness
  for two-dimensional shapes and its application in regionalization problems,''
  \emph{International Journal of Geographical Information Science}, vol.~27,
  no.~6, pp. 1227--1250, 2013.

\bibitem{polsby:91}
D.~D. Polsby and R.~D. Popper, ``The third criterion: Compactness as a
  procedural safeguard against partisan gerrymandering,'' \emph{Yale L. \&
  Pol'y Rev.}, vol.~9, p. 301, 1991.

\bibitem{schwartzberg65}
J.~E. Schwartzberg, ``Reapportionment, gerrymanders, and the notion of
  compactness,'' \emph{Minn. L. Rev.}, vol.~50, p. 443, 1965.

\bibitem{Reock:61}
\BIBentryALTinterwordspacing
E.~C. Reock, ``A note: Measuring compactness as a requirement of legislative
  apportionment,'' \emph{Midwest Journal of Political Science}, vol.~5, no.~1,
  pp. 70--74, 1961. [Online]. Available:
  \url{http://www.jstor.org/stable/2109043}
\BIBentrySTDinterwordspacing

\bibitem{triangles-de1992line}
L.~De~Floriani and E.~Puppo, ``An on-line algorithm for constrained delaunay
  triangulation,'' \emph{CVGIP: Graphical Models and Image Processing},
  vol.~54, no.~4, pp. 290--300, 1992.

\bibitem{triangles-baker1988nonobtuse}
B.~S. Baker, E.~Grosse, and C.~S. Rafferty, ``Nonobtuse triangulation of
  polygons,'' \emph{Discrete \& Computational Geometry}, vol.~3, no.~2, pp.
  147--168, 1988.

\bibitem{triangles-bern1995linear}
M.~Bern, S.~Michell, and J.~Ruppert, ``Linear-size nonobtuse triangulation of
  polygons,'' \emph{Discrete \& Computational Geometry}, vol.~14, no.~4, pp.
  411--428, 1995.

\bibitem{levcopoulos1984bounds}
C.~Levcopoulos and A.~Lingas, ``Bounds on the length of convex partitions of
  polygons,'' in \emph{International Conference on Foundations of Software
  Technology and Theoretical Computer Science}.\hskip 1em plus 0.5em minus
  0.4em\relax Springer, 1984, pp. 279--295.

\bibitem{greene1983decompositionConvex}
D.~H. Greene, ``The decomposition of polygons into convex parts,''
  \emph{Computational Geometry}, vol.~1, pp. 235--259, 1983.

\bibitem{hertel1983fast}
S.~Hertel and K.~Mehlhorn, ``Fast triangulation of simple polygons,'' in
  \emph{International Conference on Fundamentals of Computation Theory}.\hskip
  1em plus 0.5em minus 0.4em\relax Springer, 1983, pp. 207--218.

\bibitem{lien2006approximate}
J.-M. Lien and N.~M. Amato, ``Approximate convex decomposition of polygons,''
  \emph{Computational Geometry}, vol.~35, no. 1-2, pp. 100--123, 2006.

\bibitem{keil1985decomposing}
J.~M. Keil, ``Decomposing a polygon into simpler components,'' \emph{SIAM
  Journal on Computing}, vol.~14, no.~4, pp. 799--817, 1985.

\bibitem{area-constr-page1992area}
W.~Page and K.~Sastry, ``Area-bisecting polygonal paths,'' \emph{The Fibonacci
  Quarterly}, vol.~30, pp. 263--73, 1992.

\bibitem{Adjiashvili:10}
\BIBentryALTinterwordspacing
D.~Adjiashvili and D.~Peleg, ``Equal-area locus-based convex polygon
  decomposition,'' \emph{Theoretical Computer Science}, vol. 411, no.~14, pp.
  1648 -- 1667, 2010, structural Information and Communication Complexity
  (SIROCCO 2008). [Online]. Available:
  \url{http://www.sciencedirect.com/science/article/pii/S0304397510000150}
\BIBentrySTDinterwordspacing

\bibitem{berger:2016}
C.~{Berger}, M.~{Wzorek}, J.~{Kvarnstr{\"o}m}, G.~{Conte}, P.~{Doherty}, and
  A.~{Eriksson}, ``Area coverage with heterogeneous uavs using scan patterns,''
  in \emph{2016 IEEE International Symposium on Safety, Security, and Rescue
  Robotics (SSRR)}, Oct 2016, pp. 342--349.

\bibitem{Agarwal:06}
A.~{Agarwal}, {Lim Meng Hiot}, {Nguyen Trung Nghia}, and {Er Meng Joo},
  ``Parallel region coverage using multiple uavs,'' in \emph{2006 IEEE
  Aerospace Conference}, March 2006, pp. 8 pp.--.

\bibitem{agarwal2007rectilinear}
A.~Agarwal, L.~M. Hiot, E.~M. Joo, and N.~T. Nghia, ``Rectilinear workspace
  partitioning for parallel coverage using multiple unmanned aerial vehicles,''
  \emph{Advanced Robotics}, vol.~21, no. 1-2, pp. 105--120, 2007.

\bibitem{pf-robotics-Khatib1990}
\BIBentryALTinterwordspacing
O.~Khatib, \emph{Real-Time Obstacle Avoidance for Manipulators and Mobile
  Robots}.\hskip 1em plus 0.5em minus 0.4em\relax New York, NY: Springer New
  York, 1990, pp. 396--404. [Online]. Available:
  \url{https://doi.org/10.1007/978-1-4613-8997-2_29}
\BIBentrySTDinterwordspacing

\bibitem{Hansen2006}
N.~Hansen, \emph{The CMA Evolution Strategy: A Comparing Review}.\hskip 1em
  plus 0.5em minus 0.4em\relax Berlin, Heidelberg: Springer Berlin Heidelberg,
  2006, pp. 75--102.

\bibitem{BOSE2006554}
\BIBentryALTinterwordspacing
P.~Bose, S.~Cabello, O.~Cheong, J.~Gudmundsson, M.~van Kreveld, and
  B.~Speckmann, ``Area-preserving approximations of polygonal paths,''
  \emph{Journal of Discrete Algorithms}, vol.~4, no.~4, pp. 554 -- 566, 2006.
  [Online]. Available:
  \url{http://www.sciencedirect.com/science/article/pii/S157086670500050X}
\BIBentrySTDinterwordspacing

\bibitem{RAMER1972244}
\BIBentryALTinterwordspacing
U.~Ramer, ``An iterative procedure for the polygonal approximation of plane
  curves,'' \emph{Computer Graphics and Image Processing}, vol.~1, no.~3, pp.
  244 -- 256, 1972. [Online]. Available:
  \url{http://www.sciencedirect.com/science/article/pii/S0146664X72800170}
\BIBentrySTDinterwordspacing

\bibitem{doi:10.3138/FM57-6770-U75U-7727}
\BIBentryALTinterwordspacing
D.~H. DOUGLAS and T.~K. PEUCKER, ``Algorithms for the reduction of the number
  of points required to represent a digitized line or its caricature,''
  \emph{Cartographica: The International Journal for Geographic Information and
  Geovisualization}, vol.~10, no.~2, pp. 112--122, 1973. [Online]. Available:
  \url{https://doi.org/10.3138/FM57-6770-U75U-7727}
\BIBentrySTDinterwordspacing

\bibitem{gradDesc-ruder2016overview}
S.~Ruder, ``An overview of gradient descent optimization algorithms,''
  \emph{arXiv preprint arXiv:1609.04747}, 2016.

\bibitem{citeulike:3033145}
\BIBentryALTinterwordspacing
G.~Green, \emph{An Essay on the Application of mathematical Analysis to the
  theories of Electricity and Magnetism}, Nottingham, Jul. 1828. [Online].
  Available: \url{http://arxiv.org/abs/0807.0088}
\BIBentrySTDinterwordspacing

\bibitem{EFRAT2000215}
\BIBentryALTinterwordspacing
A.~Efrat, M.~J. Katz, F.~Nielsen, and M.~Sharir, ``Dynamic data structures for
  fat objects and their applications,'' \emph{Computational Geometry}, vol.~15,
  no.~4, pp. 215--227, 2000. [Online]. Available:
  \url{https://www.sciencedirect.com/science/article/pii/S0925772199000590}
\BIBentrySTDinterwordspacing

\bibitem{langford1994generating}
M.~Langford and D.~J. Unwin, ``Generating and mapping population density
  surfaces within a geographical information system,'' \emph{The Cartographic
  Journal}, vol.~31, no.~1, pp. 21--26, 1994.

\end{thebibliography}

\end{document}